\documentclass[journal]{IEEEtran}

\usepackage{cite}
\usepackage{times}
\usepackage{epsfig}
\usepackage{graphicx}
\usepackage{amsmath}
\usepackage{amssymb}
\usepackage{amsthm}
\usepackage{algorithm}
\usepackage[noend]{algpseudocode}
\usepackage{varwidth}
\usepackage{multirow}
\usepackage{color}
\usepackage{colortbl}
\usepackage{xfrac}
\definecolor{gray}{rgb}{0.5,0.5,0.5}
\usepackage{booktabs}
\usepackage{anyfontsize}
\usepackage[switch]{lineno}
\usepackage{lipsum}
\usepackage{soul}
\usepackage{float}
\usepackage{booktabs}
\usepackage[para]{threeparttable}
\usepackage{dblfloatfix}
\usepackage[breaklinks=true,bookmarks=false]{hyperref}

\DeclareMathOperator*{\argmin}{arg\,min}

\newcommand{\x}{\mathbf{x}}
\newcommand{\n}{\mathbf{n}}
\newcommand{\y}{\mathbf{y}}

\newcommand{\z}{\mathbf{z}}
\newcommand{\p}{\mathbf{p}}

\newcommand{\rr}{\mathbf{r}}
\newcommand{\sss}{\mathbf{s}}
\newcommand{\ca}[1]{\mathcal{#1}}
\newcommand{\bb}[1]{\mathbb{#1}}

\newcommand{\R}{\bb{R}}

\newcommand{\N}{\bb{N}}
\newcommand{\A}{\mathbf{A}}
\newcommand{\D}{\mathbf{D}}
\newcommand{\HH}{\mathbf{H}}

\newcommand{\I}{\bm{I}}

\newcommand{\W}{\mathbf{W}}
\newcommand{\T}{\mathbf{T}}
\newcommand{\Ss}{\bm{S}}

\newcommand{\NN}{\bm{N}}
\newcommand{\PPhi}{\mathbf{\Phi}}
\newcommand{\Vvarepsilon}{\bm{\varepsilon}}
\newcommand{\Ssigma}{\bm{\sigma}}
\newcommand{\Mmu}{\bm{\mu}}
\newcommand{\Xxi}{\bm{\xi}}
\def \<= {%
  \leq}%
\def \>= {%
  \geq}%
\newcommand{\etal}{\textit{et al}.~}
\newtheorem{thm}{Theorem}[section]
\newtheorem{prop}[thm]{Proposition}
\usepackage{mathtools,xparse}
\usepackage{amsmath,bm}
\NewDocumentCommand{\normL}{ s O{} m }{%
  \IfBooleanTF{#1}{\norm*{#3}}{\norm[#2]{#3}}_{L_2(\Omega)}%
}

\makeatletter
\def\BState{\State\hskip-\ALG@thistlm}
\makeatother

\newcommand\norm[1]{\left\lVert#1\right\rVert}

\ifCLASSINFOpdf

\else

\fi

\begin{document}

\title{Bilateral Spectrum Weighted Total Variation for Noisy-Image Super-Resolution and Image Denoising}

%\linenumbers
\author{Kaicong Sun, Sven Simon% <-this % stops a space
%\thanks{Manuscript received xxxxxxx, 2019; revised xxxxxxxxx, 2019.xxxxxxxxxxxxxxxxxxxxxxxxxxxxxxxxxxxxxyxxxxxxxxxxxxxxxxxxxx xxxxxxxxxxxxx xxxxx xxxxxxxxxxxx xxxxxxxxxxxx xxxxxxxxxxxxxxx xxxxxxxxxxxxxxxxxxxxxxxxxxxx xxxxxxxxx xxxx.}
\thanks{K. Sun and S. Simon are with the Department
of Parallel Systems, Institute of Parallel and Distributed Systems, University of Stuttgart, %Universit\"atsstra\ss e 38, Stuttgart, 70569, 
70569, Stuttgart, Germany (e-mail: kaicong.sun@ipvs.uni-stuttgart.de; sven.simon@ipvs.uni-stuttgart.de). This work was supported by the Federal Ministry of Education and Research (BMBF, Germany) under the grand No. 105M18VSA. }
%\thanks{This paper has source code available at \url{http://ieeexplore.ieee.org}. }
}
% make the title area
\maketitle

\begin{abstract}
In this paper, we propose a regularization technique for noisy-image super-resolution and image denoising. Total variation (TV) regularization is adopted in many image processing applications to preserve the local smoothness. However, TV prior is prone to oversmoothness, staircasing effect, and contrast losses. Nonlocal TV (NLTV) mitigates the contrast losses by adaptively weighting the smoothness based on the similarity measure of image patches. Although it suppresses the noise effectively in the flat regions, it might leave residual noise surrounding the edges especially when the image is not oversmoothed. To address this problem, we propose the bilateral spectrum weighted total variation (BSWTV). Specially, we apply a locally adaptive shrink coefficient to the image gradients and employ the eigenvalues of the covariance matrix of the weighted image gradients to effectively refine the weighting map and suppress the residual noise. In conjunction with the data fidelity term derived from a mixed Poisson--Gaussian noise model, the objective function is decomposed and solved by the alternating direction method of multipliers (ADMM) algorithm. In order to remove outliers and facilitate the convergence stability, the weighting map is smoothed by a Gaussian filter with an iteratively decreased kernel width and updated in a momentum-based manner in each ADMM iteration. We benchmark our method with the state-of-the-art approaches on the public real-world datasets for super-resolution and image denoising. Experiments show that the proposed method obtains outstanding performance for super-resolution and achieves promising results for denoising on real-world images. 
\end{abstract}


\IEEEpeerreviewmaketitle

\section{Introduction}
\label{introduction}
\IEEEPARstart{S}{}uper-resolution (SR) and image denoising are both fundamental and challenging image processing tasks. Super-resolution is an image enhancement technique dedicated to improving the image spatial resolution and image denoising is an image restoration task aiming to recover the underlying clean image from the noisy counterpart. In many applications, such as medical diagnostics, remote sensing, surveillance, and astronomy, due to the inherent limitation of imaging systems and imaging conditions, SR and denoising are usually required to boost the image quality and the visual perception. Generally, imaging system can be formulated as $\y = \A\x + \Vvarepsilon$ where $\A$ represents the system matrix, $\x$ denotes the latent image and $\Vvarepsilon$ indicates an additive noise. We have a multi-frame SR problem as $\A = \mathbf{DBM}$~\cite{sr3,sr4} where matrices $\mathbf{D,B,M}$ describes respectively downsampling, blurring, and motion effects and a denoising problem when $\A$ being an identity matrix. Due to their severe ill-posedness, image prior plays an essential role to regularize the solution domain. Over the last decades, fruitful study of image priors has been conducted, including approaches based on nonlocal self-similarity~\cite{NLM, buades2, BM3D}, image gradient~\cite{tv,sr0, nltv}, sparsity~\cite{sparsity1, sparsity2}. In recent years, we have witnessed the great success of deep learning and the learning-based methods~\cite{autoencoder, deepprior, SRCNN, DnCNN, DPSR} benefit significantly from the massive amount of the training data. However, the learning-based methods usually suffer from two major drawbacks. First, the performance of the deep learning approaches highly relies on the training datasets. In practice, it might be difficult to prepare synthetic datasets which adequately resemble the real-world measurements covering diverse imaging conditions. Even for unsupervised learning which do not require ground-truth, assembling enough measurements for training might also be challenging. Second, although networks are able to describe more sophisticated image priors, the functions generated by the networks are uninterpretable. Especially, the existence of undesirable artifacts is unpredictable which may encumber the employment of the learning-based approaches in applications such as geometric dimensioning and non-destructive testing.

In this work, inspired by the nonlocal TV (NLTV)~\cite{nltv}, we intend to smooth the image in the flat regions and meanwhile maintain the sharpness at edges by applying a locally adaptive weighting map to the total variation. Different from the NLTV, our bilateral spectrum weighted total variation (BSWTV) shrinks the mask of the edges in the weighting map explicitly such that it suppresses the residual noise surrounding the edges effectively without compromising the sharpness. The contribution of this work is summarized as follows: 
\begin{itemize}
\item We propose a regularization technique for effectively suppressing the noise and meanwhile preserving the sharpness of fine structures. Particularly, a shrink coefficient is introduced to adaptively weight the image gradient and the weighting map of TV is estimated based on the spectrum of the weighted-gradient covariance matrix.
\item Combining the proposed BSWTV with the data fidelity term derived from a mixed Poisson-Gaussian noise model, the overall objective function is solved based on the ADMM algorithm where the update of the weighting map is integrated as the first step in the ADMM framework. In order to remove outliers and facilitate the convergence stability, the weighting map is smoothed by a Gaussian filter with an iteratively decreased kernel width and updated in a momentum-based fashion.
\item We benchmark our approach with the state-of-the-art methods on the public real-world datasets for super-resolution and image denoising. The proposed method shows promising performance on the real-world images.

\end{itemize}

\section{Related Work}
\label{ssec:Prior}
In last decades, there has been intensive investigation on SR~\cite{sr0, singh, Yuan, sun, kohler, SRCNN, VDSR, EDSR, DPSR, ren} and image denoising~\cite{tv, NLM, buades2, nltv, zhangNLTV, BM3D, EPLL, NCSR, WNNM, DnCNN,lefkimmiatis}. Due to the ill-posedness of the problems, the existing methods employ either explicitly handcrafted image priors or implicit priors. Specially, the majority of the optimization-based traditional methods exploit the handcrafted priors. Rudin~\etal\cite{tv} introduce the total variation (TV) as the regularization for image denoising. In~\cite{sr0}, bilateral total variation (BTV) is proposed by concerning photometric and geometric distance in an extended neighborhood for multi-frame SR. Yuan~\etal\cite{Yuan} propose a regional spatially adaptive total variation (RSATV) for SR based on spatial information filtering and clustering which partition the image into multiple segments. However, pixels within each segment are limited to an equal weight. 

Besides the above gradient-based priors, nonlocal-means (NL-means)~\cite{NLM} based on the self-similarity exploits the natural redundancy of image patterns aiming to average the pixels which are surrounded by similar textures. Specially, the NL-means algorithm is formulated as
\begin{linenomath}
\begin{equation}
x(i) =\sum\limits_{j\in R_i}w(i,j)x(j),
\end{equation}
\end{linenomath}
where $x(i)$ is the estimated image pixel. Weight $w(i,j)$ depicts the similarity between pixel $i$ and $j$ with $\sum_jw(i,j)=1, j\in R_i$ where $R_i$ denotes the search window of pixel $i$. The weight $w(i,j)$ is measured by
\begin{linenomath}
\begin{equation}
w(i,j) =\frac{1}{Z(i)}exp(-\frac{||\NN(i)-\NN(j)||^2_{2,\sigma}}{\eta^2}).
\end{equation}
\end{linenomath}
$\NN(i), \NN(j)$ indicate respectively the square neighborhood of pixel $i$ and $j$. $||\NN(i)-\NN(j)||^2_{2,\sigma}$ is the weighted Euclidean distance with $\sigma$ being the standard deviation of the Gaussian kernel. $Z(i)$ denotes the normalizing constant with $Z(i)=\sum_j exp(-\frac{||\NN(i)-\NN(j)||^2_{2,\sigma}}{\eta^2})$ and $\eta$ being the constant filtering parameter. In~\cite{adaptivenlm}, the authors combine the NL-means with the TV regularization with the purpose of exploring both the redundant and the smooth properties of images to overcome the drawbacks of oversmoothness and inefficient denoising. BM3D~\cite{BM3D} is another discriminative denoising method based on self-similarity which groups similar patches and performs collaborative filtering by shrinkage in the 3D transform domain. 

To overcome the performance decline caused by contrast losses, Gilboa~\etal\cite{nltv} propose a variational regularization,  nonlocal TV (NLTV), based on the self-similarity which can be formulated as
\begin{linenomath}
\begin{equation}\label{nltv}
V_c(\x)= \sum\limits_{\D=(D_x, D_y)}||\PPhi_D(\Ss_D-\I)\x||_1,
\end{equation}
\end{linenomath}
where $(D_x, D_y)$ indicates the shift vector with $D_x, D_y\in[-(R-1)/2, (R-1)/2]$ and $R$ is the window size. Matrix $\Ss_D$ acts as the shift operator and $\PPhi_D$ represents the weighting map associated with the shift vector $\D$ defined as 
\begin{linenomath}
\begin{equation}\label{Phi}
\PPhi_D(i,j)= exp(-\frac{||\NN(i,j)-\NN(i+D_x, j+D_y)||^2_2}{\eta^2}).
\end{equation}
\end{linenomath}
$\NN(i,j)$ denotes the neighbors of the center pixel $(i,j)$ in the similarity patch of size $r$ and $\eta$ is the filtering parameter which controls the smoothness. An advantage of NLTV over NL-means is that we can easily plug the regularization technique into different image processing tasks. However, NLTV usually suffers from the drawback that it is prone to residual noise in the surroundings of the edges especially when the image is not oversmoothed.

In recent years, learning-based methods have achieved great success in many applications. Most of the work takes advantage of learning features and implicit image priors from external datasets~\cite{dic0, dic1, DnCNN, lefkimmiatis, dic3, SRCNN, VDSR, EDSR, DPSR, ren}. Particularly, Dong~\etal\cite{SRCNN} introduce a convolutional neural network (CNN) for single-frame SR. Afterwards, a series of work~\cite{VDSR, EDSR, DPSR} has achieved noticeable performance. Lim~\etal\cite{EDSR} propose a deep and compact residual network EDSR by effectively removing unnecessary modules in the conventional residual networks. In~\cite{bao}, the authors propose an iterative network structure for super-resolving noisy images contaminated by an additive white Gaussian noise (AWGN). In the literature of SR and imaging denoising, most of the models are derived based on the assumption that the images are corrupted by the AWGN. However, in reality, the composition of noise in imaging systems is more sophisticated. There are mainly two sources of noise dominating in digital imaging process: the intensity-independent readout noise and reset noise which can be modeled as an additive Gaussian noise and the intensity-dependent photon shot noise which obeys a Poisson distribution~\cite{Snyder, Aguerrebere2}. To the best of our knowledge, despite the importance of adopting an accurate noise description in the imaging model, the literature on SR and image denoising based on a mixed Poisson--Gaussian noise model is limited~\cite{Luisier, Zhang, Yann, sun}.

In this paper, we address the drawback of the NLTV and propose a generalized algorithm coping with mixed Poisson-Gaussian noise for real-world noisy-image super-resolution and image denoising. %Specially, we introduce a novel variational regularization approach which leverages the locally adaptive weights derived from the spectrum of the weighted-gradient covariance matrix. Combining with the data fidelity term derived from a mixed Poisson--Gaussian noise model in~\cite{sun}, the overall objective function is decomposed into subproblems and solved by the ADMM algorithm.  

\section{Methods}
\label{sec: Methods}
\subsection{Bilateral Spectrum Weighted Total Variation}

TV performs smoothing based on the image gradients without concerning the features such that it is prone to oversmoothness, staircasing effect, and contrast losses. In order to alleviate the contrast losses, inspired by~\cite{kernelregression}, we employ the information entailed in the gradient covariance matrix to locate features such as edges. Generally, flat regions and edges can be distinguished by the
spectrum of the gradient covariance matrix~\cite{harris}. For images contaminated by a mixed Poisson--Gaussian noise, according to Proposition~\ref{thm:GaussianDist}, no matter if it has a white Gaussian component, we can formulate the gradients in the flat regions by a white isotropic Gaussian distribution with a limited variance which theoretically enables us to differentiate the mixed noise in flat areas from the edges. 

\begin{prop}\label{thm:GaussianDist}
Let us define an observed digital image $y: \N^2_0\to \R$ contaminated by a mixed Poisson--Gaussian noise as $y = z + n_p(z) + n_g$ where $(z_i+n_p(z_i))/\alpha\sim P(z_i/\alpha)$ with $z_i$ being the expected pixel value and $\alpha$ being a scalar. $n_g$ is an independent additive Gaussian noise with $n_g(i)\sim N(\mu_{i},\sigma^2_{i})$. If we have a homogeneous region $\Omega\subset \N^2_0$, where $\forall i, j \in \Omega$, $|z_i+\mu_i-z_j-\mu_j|<\varepsilon_1, |\alpha z_i+\sigma^2_i-\alpha z_j-\sigma^2_j|<\varepsilon_2,  \forall \varepsilon_1, \varepsilon_2>0$, then the gradient of each element $i$ in $\Omega$ has the same isotropic white Gaussian distribution $\nabla_xy(i), \nabla_yy(i)\sim N(0, (\alpha z_i+\sigma^2_i)/2)$ and a collection of the gradients obeys an isotropic white Gaussian distribution. (See Appendix~\ref{App:proof} for a proof)
\end{prop}

\begin{figure}
	\centering
	 \includegraphics[width=0.45\textwidth]{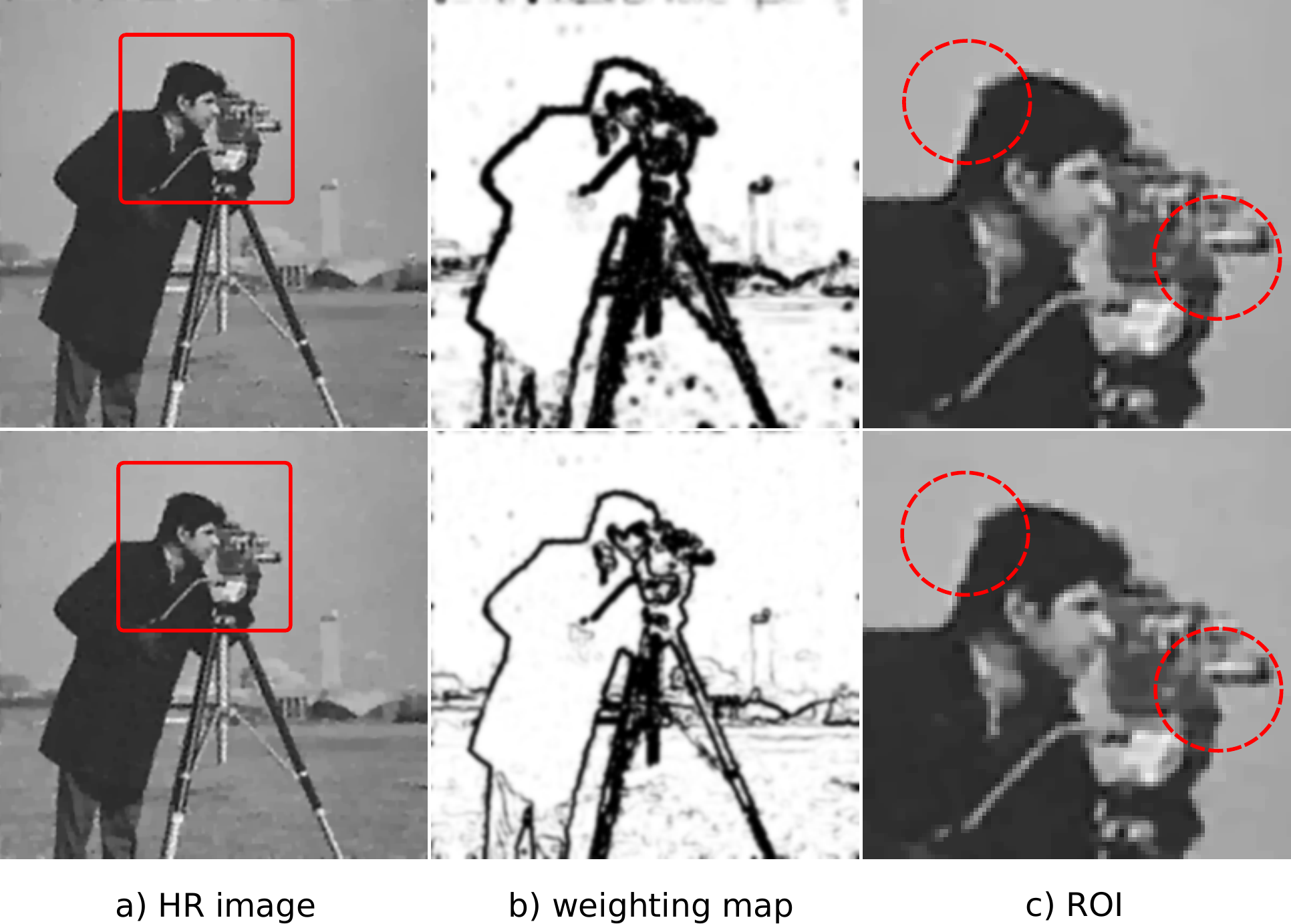}		
	\caption{Impact of the decay parameter $\gamma$ on the SR performance (2$\times$). Top: $\gamma = 1$, PSNR = 30.35dB, SSIM = 0.8577; Bottom: $\gamma = 0.8$, PSNR = 30.47dB, SSIM = 0.8607.}
	\label{fig:camera}
\end{figure}

In order to efficiently suppress the noise in flat regions and meanwhile preserve the fine structures, we propose a novel regularizer BSWTV as formulated in Eq.~\eqref{BSWTV}. BSWTV possesses the merit that it enables a refinement of the weighting map by introducing an inhomogeneous shrink coefficient. 
\begin{equation}\label{BSWTV}
BSWTV(\x) := ||\PPhi\nabla \x||_1, \PPhi =diag[\phi_1,\cdots, \phi_n]
\end{equation}
The $i$th diagonal element of the weighting map $\PPhi$ is defined as
\begin{equation}\label{weight}
\phi_i = exp(-|\lambda_{i1}-\lambda_{i2}|/\eta^2),
\end{equation}
where $\eta$ is the smoothing parameter which controls the dynamic range of $\PPhi$ and $\lambda_{i1}, \lambda_{i2}$ are the eigenvalues of the covariance matrix of the bilateral weighted gradients $G_i$ which is formulated as

\begin{equation}\label{gradient}
\begin{split}
&G_i =
\begin{bmatrix}
\omega_1g^1_{x}, \dots, \omega_jg^j_{x}, \dots, \omega_qg^q_{x}\\
\omega_1g^1_{y}, \dots, \omega_jg^j_{y}, \dots, \omega_qg^q_{y}\\
\end{bmatrix},  ~\omega_j = \xi_j^{d(i,j)} \\ &\textit{with } \xi^k_j =\xi^{k-1}_j(\gamma+\frac{(1-\gamma)}{1+exp^{f(\PPhi^{k-1}_{N_j})}}), ~ j\in \NN_i.
\end{split}
\end{equation}
$g_j$ represents the gradient at pixel $j$ and is expressed as $g_j:=(g^j_{x}, g^j_{y}) = (\nabla_xx_j, \nabla_yx_j)$. The square patch centered at pixel $i$ has the amount of $q = r^2$ pixels and is defined as $\NN_i =\{j: |i-j|\leq (r-1)/2\}$ with $r$ being the odd size of the patch. $\omega_j$ acts as the weight assigned to each individual neighbor in $\NN_i$ and indicates the significance of the neighbor $j$ to the center pixel $i$ which depends on the distance $d(i,j):= |d_x(i,j)|+|d_y(i,j)|$ along x and y axis with $d_x, d_y\in[-(r-1)/2, (r-1)/2]$ and the local adaptive shrink coefficient $\xi_j$. The superscript of $\xi_j^{k}$ depicts the $k$th iteration of the ADMM algorithm and the decay scalar $\gamma\in[0,1]$ serves for the shrinkage of the spread of the gradients within the patch. The intuition of introducing $\xi$ and $\gamma$ is to adaptively ``squeeze'' the gradient matrix $G$ such that the discrepancy between eigenvalues $\lambda_{i1}$ and $\lambda_{i2}$ decreases as the algorithm converges and the mask of the edges in the weighting map $\PPhi$ becomes thinned. $\PPhi_{N_j}$ denotes the neighbors of pixel $j$ in the weighting map $\PPhi$. $f$ is a function of $\PPhi_{N_j}$ which controls the whitening based on the image content and enables an inhomogeneous decay of $\Xxi$. Specially, for flat regions, $f$ is supposed to be a large positive value such that the shrink coefficient $\xi_j$ is decreased by $\gamma$ and the weighting map gets further whitened, while for fine structures, $f$ should be a large negative value so that the shrink coefficient $\xi_j$ is not attenuated. A simple choice of $f(\x)$ could be an affine function $f(\x):=a (\bar{\x}-b)$ where $\bar{\x}$ denotes the mean of vector $\x$ and the positive scalars $a, b$ are the amplitude and shift parameters, respectively. Hence, based on the previous $\PPhi$, map $\Xxi$ is inhomogeneously shrinked by factors in the range of $(\gamma, 1)$. In Fig.~\ref{fig:camera}, we illustrate the effectiveness of leveraging the decay parameter $\gamma$. The top row illustrates the weighting map $\PPhi$ and the SR image without decaying the shrink coefficient. The bottom row exhibits the results with $\gamma = 0.8$. As we can see, the weighting map $\PPhi$ in the bottom row has much thinned mask for edges than the counterpart in the top row under the same smoothing parameter $\eta$. Consequently, the SR image has much cleaner and pleasant contours without oversmoothing the fine structures. We demonstrate a detailed analysis of the effectiveness of the decay scalar $\gamma$, the smoothing parameter $\eta$, and the shift parameter $b$ on the reconstruction performance in Section~\ref{parameter analysis}. %The top row illustrates the weighting map and the SR image by $b = 0$ and the bottom row shows the results by $b = 1$. As shown, more detailed structures  can be masked out in the weighting map and hence preserved in the SR image. 

The update of the weighting map $\PPhi$ is embedded in the ADMM framework as described in Algorithm~\ref{pseudocode MPSR} in Section~\ref{sec:update overall}. For the sake of suppressing the outliers and enhancing the convergence stability in the ADMM update scheme, the weighting map $\PPhi$ calculated by Eqs.~\eqref{weight} and~\eqref{gradient} is followed by a Gaussian filter with an iteratively decreased kernel width and updated in a momentum-based fashion in each ADMM iteration. A detailed description of the update of the weighting map $\PPhi$ is given in Section~\ref{sec:update weighting map}.

\iffalse
\begin{figure}
	\centering
	 \includegraphics[width=0.45\textwidth]{Fig2.pdf}		
	\caption{Impact of the inhomogeneous decay by the shift parameter $b$ in function $f$ on the SR performance (2$\times$). Top: $b = 0$, PSNR = 29.96dB, SSIM = 0.8011; Bottom: $b = 1$, PSNR = 30.03dB, SSIM = 0.8261.}
	\label{fig:lena}
\end{figure}
\fi

\subsection{Super-Resolution and Image Denoising Based on BSWTV}
In digital imaging systems as mentioned in Section~\ref{ssec:Prior}, the intensity-dependent photon shot noise arises from the stochastic nature of the photon-counting process and can be modeled as a Poisson noise. Meanwhile, Gaussian noise exists due to the intrinsic thermal and electronic fluctuations in the sensors~\cite{Luisier}. Therefore, the imaging system is formulated based on a mixed Poisson-Gaussian noise model as following:
\begin{linenomath}
\begin{equation}\label{system}
y_i = z_i + n_p(z_i) + n_g,
\end{equation}
\end{linenomath}
where $y_i$ stands for the intensity value at the $i$th pixel of the observed image $\y$ which is contaminated by a mixed Poisson--Gaussian noise. $z_i$ indicates the clean pixel value. $n_p(z_i)$ is an intensity-dependent noise with $(z_i+n_p(z_i))/\alpha\sim P(z_i/\alpha)$ where $\alpha$ is a scalar accounting for quantum efficiency and analog gain~\cite{Egiazarian}. $n_g$ represents an additive Gaussian noise with $n_g\sim N(\mu_i, \sigma^2_i)$. $n_p(z_i)$ describes mostly photon shot noise and $n_g$ embodies mainly readout noise and reset noise. In fact, there are other noises existing in the complementary metal-oxide-semiconductor (CMOS) or charge-coupled device (CCD) detectors such as the Poissonian dark current shot noise which is negligible for exposure time less than 1$s$ and the quantization noise which is uniformly distributed and can be omitted compared to the readout noise except in very low-illumination conditions~\cite{Aguerrebere2}. Without loss of generality, we define $\mathbf{A=DBM}$ being the system matrix as described in Section~\ref{introduction} and $\x^{\ast} = [x_1^{\ast},\dots, x_N^{\ast}]$ being the vectorized latent image with $\z = \A\x^{\ast}$.

Assuming that $n_p$ and $n_g$ are mutually independent, we yield the mean and the variance of the intensity of pixel $i$ as
\begin{linenomath}
\begin{equation}\label{meanvar}
\begin{split}
\mathit{E}(y_i) &= \mathit{E}(z_i + n_p)+\mathit{E}(n_g) = [\A]_i\x^{\ast} + \mu_i\\
\mathit{Var}(y_i) &= \mathit{Var}(z_i + n_p)+\mathit{Var}(n_g) = \alpha[\A]_i\x^{\ast}+\sigma^2_i,
\end{split}
\end{equation}
\end{linenomath}
where $[\A]_i$ indicates the $i$th row of matrix $\A$ and $\x^{\ast}$ represents the expected image. It should be noted that in this paper, the degradation matrix $\A$, the scalar $\alpha$ and the Gaussian noise parameters $\mu_i$ and $\sigma_i$ are assumed to be known. According to the Central Limit Theorem (CLT), we have $P(z_i/\alpha)\simeq N(z_i/\alpha,z_i/\alpha)$ as  $z_i/\alpha$ being sufficiently large. Hence, the observed value $y_i$ can be approximated by a Gaussian distribution. Based on Eq.~\eqref{meanvar}, we yield $y_i\sim  \mathcal{N}([\A]_i\x^{\ast} + \mu_i, \alpha[\A]_i\x^{\ast}+\sigma^2_i)$. %when $[\A]_i\x^{\ast}$ is sufficiently large. 
Therefore, the probability mass function (PMF) of $y_i$ conditioned on the expected image $\x^{\ast}$ is expressed as 
\begin{linenomath}
\begin{equation}\label{gaussian}
\begin{split}
P(y_i|\x^{\ast}) = \frac{1}{\sqrt{2\pi (\alpha[\A]_i\x^{\ast}+\sigma_i^2)}}\exp{\frac{-(y_i-[\A]_i\x^{\ast}-\mu_i)^2}{2(\alpha[\A]_i\x^{\ast}+\sigma_i^2)}}.
\end{split}
\end{equation}
\end{linenomath}

According to Bayes' theorem, we have the associated negative log-likelihood formulated as
\begin{linenomath}
\begin{equation}\label{Log}
\begin{split}
&-\log{P(\y|\x)}= -\log{\prod\limits_{i=1}^{n}P(y_i|\x)}\\
&=\frac{1}{2}\left(\parallel{\y-\A\x-\Mmu}\parallel_{\W}^{2}+\langle\log{(\alpha\A\x+\Ssigma^{2})},1\rangle\right) + c,
\end{split}
\end{equation}
\end{linenomath}
where log$(\cdot)$ is an elementwise operation, $\langle\cdot,\cdot\rangle$ indicates the inner product and \textit{c} is a constant. For the sake of brevity, we will omit the constant \textit{c} in the latter formulation. The intensity-dependent diagonal weight matrix $\W$ is expressed as
\begin{linenomath}
\begin{equation}
\W = \text{diag}\{\frac{1}{\alpha[\A]_i\x+\sigma^2_i}\}.
\end{equation}
\end{linenomath}
Specially, for image denoising we set $\A = \I$ and for single-frame SR, we can formulate $\mathbf{A = DB}$. With regard to multi-frame SR, instead of having one observed low-resolution (LR) image $\y$, there are $m$ LR images $\y_i$ with the individual system matrix $\A_i$ and additive noise $\n_i$. Assuming the LR images are independent, we can extend the formulation for single-frame input expressed in Eq.~\eqref{Log} as below:   
\begin{linenomath}
\begin{equation}\label{dataterm}
\begin{split}
&-\log{P(\y_1, \dots, \y_m|\x)} = -\sum\limits_{i=1}^{m}\log{P(\y_i|\x)}\\
&= \frac{1}{2}\sum\limits_{i=1}^{m}\left(\parallel{\y_i-\A_i\x-\Mmu_i}\parallel_{\W_i}^{2}+\langle\log{(\alpha_i\A_i\x+\Ssigma_i^{2})},1\rangle\right).
\end{split}
\end{equation}
\end{linenomath}
In the rest of the paper, we formulate the data fidelity term in general by Eq.~\eqref{dataterm}. Denoising and single-frame SR are considered as the special cases with $m=1$.

Combining the regularization term expressed in Eq.~\eqref{BSWTV} with the data fidelity term formulated in Eq.~\eqref{dataterm}, we yield the overall objective function as

\begin{linenomath}
\begin{equation}
\label{energy function}
\begin{split}
\textit{J} &= \frac{1}{2}\sum\limits_{i=1}^{m}\left(\parallel{\y_i-\A_i\x-\Mmu_i}\parallel_{\W_i}^{2}+\langle\log{(\alpha_i\A_i\x+\Ssigma^2_i)},1\rangle\right)\\
&\quad + \lambda ||\PPhi\nabla \x||_1,
\end{split}
\end{equation}
\end{linenomath}
with $\lambda$ being the weight of the regularization term. Due to the fact that the data fidelity term is derived from a mixed Poisson--Gaussian noise model, we name the above algorithm as MPG+BSWTV.

\section{Optimization Method}
\label{Optimization Method}
\subsection{Decomposition and ADMM}
Considering the complexity of the algorithm, the objective function $J$ in Eq.~\eqref{energy function} can be decomposed into subfunctions such that the reformulated optimization problem can be attacked by means of constrained optimization, e.g., dual ascent and ADMM. Dual ascent is based on the Lagrangian and usually has inferior convergence properties, while ADMM benefits from the augmented Lagrangian and improves the convergence~\cite{ADMM}. In this paper, we utilize the ADMM algorithm to solve the decomposed objective function. Particularly, we adopt the anisotropic TV in the implementation: $||\PPhi\nabla \x||_1 = ||\PPhi(\Ss_x-\I)\x||_1 + ||\Phi(\Ss_y-\I)\x||_1$ where matrices $\Ss_x, \Ss_y$ perform respectively the shift operation along x and y axis by one pixel. Therefore, the overall optimization problem can be split into $m+2$ subproblems with the corresponding constraints as
\begin{linenomath}
\begin{equation}
\label{consensus}
  \begin{split}
    \argmin_{\x, \z_i} J &= \sum_{i=1}^{m+2} g_i(\z_i)\\
    \text{ subject to } \T_i\x-\z_i&= 0,\quad \forall i \in [1, m+2]
  \end{split}
\end{equation}
\end{linenomath}
with $\z_i \in \mathbb{R}^N$
and $\T_i$
being a matrix:
\begin{linenomath}
 \begin{equation} \T_i=
     \begin{cases}
     \I_{N \times N}, &i\in [1, m], \\
     \PPhi (\Ss_x-\I_{N\times N}), &i=m+1,\\
     \PPhi (\Ss_y-\I_{N\times N}), &i=m+2.
     \end{cases}
 \end{equation}
\end{linenomath}
 
Specially, $g_i(\cdot)$ is defined as following:
\begin{linenomath}
\begin{equation}
    \begin{split}
    &g_i(\z_i) := \dfrac{1}{2}\left(||\y_i - \A_i \z_i-\Mmu_i ||_{\W_i}^2  +  \langle \log (\alpha_i\A_{i}\z_i + \Ssigma_i^2 ), 1\rangle\right), \\
    &i\in[1,m],\\
    &g_i(\z_i):= \lambda||\z_i||_{1}, \ i\in[m+1, m+2].
    \end{split}
\end{equation}
\end{linenomath}

The augmented Lagrangian is hence formulated as
\begin{linenomath}
\begin{equation}
\begin{split}
  &\ca{L}_{H} (\x,\z,\p)= \sum_{i=1}^{m+2} \ca{L}_{H_i} (\x,\z_i,\p_i)\\
  &=\sum_{i=1}^{m+2}\Bigl(g_i(\z_i) + \langle \p_i,\T_i \x- \z_i\rangle +\dfrac{1}{2}||\T_i \x- \z_i||_{\HH_{i}}^2\Bigr),
\end{split}
\end{equation}
\end{linenomath}
where $\p_i$ is the dual variable associated with the individual constraint and matrix $\HH_{i}$ is defined as
\begin{linenomath}
   \begin{equation}
   \HH_{i}:=\mbox{diag}[\rho_i,\dotso ,\rho_i],\quad \forall i\in[1,\cdots, m+2]
\end{equation}
\end{linenomath}
with $\rho_{i}$ being a positive scalar acting as the update step size of the dual variable $\p_i$.

The decomposed objective function formulated in Eq.~\eqref{consensus} is solved in the following iterative scheme:

\begin{linenomath}
\begin{subequations} 
\begin{align}
    &\x^{k+1} = \argmin_{\x}\sum_{i=1}^{m+2}\dfrac{\rho_i}{2} ||\T_i \x -  \z_i^{k} + \dfrac{\p_i^k}{\rho_i} ||_2^2 \label{xupdate}\\
    &\z_i^{k+1} = \argmin_{\z_i}  g_i(\z_i) +  \dfrac{\rho_i}{2}
    ||\z_i-\T_i \x^{k+1} - \dfrac{\p_i^k}{\rho_i}||_2^2 \label{zupdate}\\
   &\p_i^{k+1} = \p_i^k + \rho_i (\T_i \x ^{k+1} - \z_i^{k+1}) \label{pupdate}.
\end{align}
\end{subequations}
\end{linenomath}

Since Eq.~\eqref{xupdate} is quadratic and differentiable, the update of $\x^{k+1}$ can be achieved by, e.g., the conjugate gradient (CG) algorithm. $g_i(\z_i)$ with $i\in\left[1, m\right]$ is nonconvex which is solved by the scaled conjugate gradient (SCG). To update $\z_i^{k+1}$ associated with the BSWTV prior, i.e., $i\in[m+1, m+2]$, we utilize the proximal operator of the L1-norm:

\begin{linenomath}
\begin{equation}
%\nonumber
\label{zproximal}
\begin{split}
   \z_i^{k+1} &= \argmin_{\z_i} \lambda||\z_{i}||_{1} + \dfrac{\rho_i}{2}
    ||\z_{i}-\T_i\x^k - \dfrac{\p_i^k}{\rho_i}||_2^2 \\&= prox_{ \lambda(
      \rho_i)^{-1}|| \cdot||_1 } (\T_i\x^k + \dfrac{\p_i^k}{\rho_i})
\end{split}
\end{equation}
\end{linenomath}

and we can yield the closed-form solution as
\begin{linenomath}
\begin{equation}
%\nonumber
\label{zproximal1}
\begin{split}
%\small
[\z_i^{k+1}]_j=
\begin{cases}
[\T_i\x^k + \dfrac{\p^k_i}{\rho_i}]_j- \dfrac{\lambda}{\rho_i} ,  &[\T_i\x^k +
\dfrac{\p_i^k}{\rho_i}]_j \>=  \dfrac{\lambda}{\rho_i}, 
\\
0,  &|[\T_i\x^k +
\dfrac{\p_i^k}{\rho_i}]_j| \<=  \dfrac{\lambda}{\rho_i}, \\
[\T_i\x^k + \dfrac{\p_i^k}{\rho_i}]_j + \dfrac{\lambda}{\rho_i} ,  &[\T_i\x^k +\dfrac{\p_i^k}{\rho_i}]_j \<= - \dfrac{\lambda}{\rho_i}.
\end{cases}
\end{split}
\end{equation}
\end{linenomath}

In order to improve the convergence and reduce the dependency of the initialization in practice, the penalty parameter $\rho_i$ is updated iteratively, for instance, by means of synchronizing the convergence of the primal residual $\rr_i^k$ and the dual residual $\sss_i^k$ with the scheme in~\cite{ADMM}:
\begin{linenomath}
\begin{equation}
\label{updaterho}
\begin{split}
%\small
\rho_i^{k+1}=
\begin{cases}
c_1\rho_i^k ,  &||\rr_i^k||_2 > c||\sss_i^k||_2, \\
\rho_i^k/c_2,  &||\sss_i^k||_2 > c||\rr_i^k||_2, \\
\rho_i^k ,  &otherwise,
\end{cases}
\end{split}
\end{equation}
\end{linenomath}
where $c_1, c_2, c$ are constants with $c_1>1, c_2>1, c>1$. The primal and dual residuals $\rr_i^k$, $\sss_i^k$ are calculated as
\begin{linenomath}
\begin{equation}
\label{residual}
\begin{split}
    \rr_i^{k+1} &= \T_i\x^{k+1}-\z_i^{k+1}\\
	\sss_i^{k+1} &= -\rho_i^k\T_i^T(\z_i^{k+1}-\z_i^k).
\end{split}
\end{equation}
\end{linenomath}

An early stopping criteria based on the primal and dual residuals is used as depicted in Algorithm~\ref{pseudocode MPSR}.

\subsection{Update of Weighting Map}\label{sec:update weighting map}

The weighting map $\PPhi$ is updated iteratively within the ADMM framework. In particular, in order to enhance the convergence stability and update efficiency, we perform two additional steps following Eqs.~\eqref{weight} and~\eqref{gradient}. Firstly, we smooth the weighting map $\PPhi$ by convolving with an isotropic 2D Gaussian kernel $G(\sigma_{\Phi})$ to alleviate the effect of the outliers on the weighting map. Secondly, we update the smoothed weighting map in a momentum-based manner to avoid strong fluctuation in the objective function during convergence. Specially, the decay scalar $\gamma$ is employed to iteratively decrease the Gaussian parameter $\sigma_{\Phi}$ and the momentum coefficient $\beta$. The insight behind decaying the width of the Gaussian kernel is to silently remove the outliers in the weighting map by convolving with a relatively wide Gaussian kernel in the first ADMM iterations and along with the convergence of the objective function, the weighting map becomes less smoothed by leveraging a narrowed Gaussian filter. Consequently, the mask of the edges in the weighting map is thinned and sharpened in a moderate manner and the remaining noise surrounding the edges is effectively suppressed. The update framework of the weighting map $\PPhi$ in the $k$th ADMM iteration is formulated as following in Algorithm~\ref{pseudocode BSWTV}.

\begin{algorithm}
\caption{Update of Weighting Map}\label{pseudocode BSWTV}
\begin{algorithmic}[1]
\State \text{Initialize $\PPhi, \Xxi, \gamma, \eta, \beta, r, \sigma_{\Phi}, \sigma_{\text{min}}$.}
\Procedure{Calculating Weighting Map}{}
\State \text{Calculate} $\Xxi^k(\Xxi^{k-1}, \PPhi^{k-1}, \gamma)$
\Comment{by Eq.~\eqref{gradient}}
\State  $\PPhi^{k}\leftarrow \text{Calculate}~\PPhi (\x^{k-1}, \Xxi^k, \eta)$ \Comment{by Eq.~\eqref{weight}}
\State $\sigma^{k}_{\Phi} = max(\sigma_{\text{min}}, \gamma \sigma^{k-1}_{\Phi})$ 
\State $\PPhi^{k} = G(\sigma^{k}_{\Phi})*\PPhi^{k}$
\State $\beta^{k} = \gamma \beta^{k-1}$. 
\State $\PPhi^{k} = \beta^{k} \PPhi^{k-1} + (1-\beta^{k}) \PPhi^{k}$
\EndProcedure
\end{algorithmic}
\end{algorithm}

\subsection{Overall Optimization Framework}\label{sec:update overall}
The core of the overall optimization is to integrate the update of the weighting map as described in Algorithm~\ref{pseudocode BSWTV} into the ADMM framework. As the weighting map $\PPhi$ is coupled with the latent image $\x$ as expressed in Eq.~\eqref{xupdate} and is used for the update of the variables $\z_i$ and $\p_i$ for $i\in[m+1, m+2]$ as formulated in Eqs.~\eqref{zproximal},~\eqref{pupdate}, we update $\PPhi$ with $\x$ fixed and perform the update of $\PPhi$ as the first step in the ADMM iteration. The pseudocode for solving the overall objective function is presented in Algorithm~\ref{pseudocode MPSR}. As depicted, in each ADMM iteration there are four main steps in sequence to respectively update $\PPhi, \x, \z$, and $\p$. Specially, the computational complexity for updating the weighting map $\PPhi$ in each ADMM iteration is $\mathcal{O}(r^2w^2N)$ with $r^2$ being the patch size and $w^2$ being the Gaussian kernel size. $\x$ is iteratively solved by CG which has a computational complexity of $\mathcal{O}(k_xN^2)$ with $k_x$ being the amount of iterations of CG. For solving $\z_i$ with respect to the data term with $i\in[1,m]$, we employ the SCG which has the computational cost of $\mathcal{O}(mk_zN^2)$ with $k_z$ being the number of iterations for SCG. For $\z_i$ associated with the regularization term, we employ the proximal operator which has the complexity of $\mathcal{O}(N^2)$. The last step is to solve the dual variables $\p$. For $\p_i$ with $i\in[1,m]$, $\T_i$ denotes the identity matrix and the computational cost is $\mathcal{O}(N)$ and for $i\in[m+1,m+2]$, the update of $\p_i$ has $\mathcal{O}(N^2)$ computational complexity. Hence, the overall computational complexity for each ADMM iteration is $\mathcal{O}((k_x+mk_z)N^2)$. %The source code is available at \url{http://ieeexplore.ieee.org}. 

\begin{algorithm}
\caption{Proposed Algorithm}\label{pseudocode MPSR}
\begin{algorithmic}[1]
\State \text{Initialize $\PPhi, \Xxi, \Ssigma, \lambda, r, \eta, \beta, \gamma, \sigma_{\Phi}, \sigma_{min}, \rho, iter, \alpha, c, c_1, c_2$}
%\State \text{Calculate system matrix $A$ based on~\ref{DBG matrix}.}
\State \text{Load observed images $\y_i\ i \in\left[1,\cdots,m\right]$}
\Procedure{Solving ADMM}{}
\State $\z_i^0 = \y_1$ for denoising, $\textit{Bicubic}(\y_1)$ for SR, $\ i
  \in\left[1,m\right]$
\State $\z_i^0 = 0,\ i
  \in\left[m+1,m+2\right]$
\State $\p_i^0 = 0,\ i
  \in\left[1,m+2\right]$
\State $\x^0 = 0$
\While{$ k < iter$}
\State \text{Update weighting map $\PPhi$}
\Comment{by Alg.~\ref{pseudocode BSWTV}}
\State \text{CG($\x^k$)}
\Comment{by Eq.~\eqref{xupdate}}
\For  {$i=1$ to $m+2$} 
\If {$i \in [1,m]$}
\State \text{SCG($\z_i^k$)}
\Comment{by Eq.~\eqref{zupdate}}
\Else
\State \text{Prox($\z_i^k)$}
\Comment{by Eqs.~\eqref{zupdate},~\eqref{zproximal},~\eqref{zproximal1}}
\EndIf
\State \text{Update $\rho_i^k$}
\Comment{by Eqs.~\eqref{updaterho},~\eqref{residual}}
\State \text{Update $\p_i^k$}
\Comment{by Eq.~\eqref{pupdate}}
\EndFor
\If {$\sum_i(||\rr_i^{k-1}||_2^2-||\rr_i^k||_2^2)/\sum_i(||\rr_i^{k-1}||_2^2) < \epsilon_1$ and $\sum_i(||\sss_i^{k-1}||_2^2-||\sss_i^k||_2^2)/\sum_i(||\sss_i^{k-1}||_2^2) < \epsilon_2$}
\State \textbf{break}
\EndIf
\State $k= k+1$
\EndWhile
\State \textbf{end while}
\State \textbf{return} \text{reconstructed image $\x$.}
\EndProcedure
\end{algorithmic}
\end{algorithm}

\section{Experiments and Results}
\label{Experiments and Results}
In this section, we conducted extensive experiments to evaluate our proposed method on the synthetic and the real-world images for both multi-frame SR and image denoising. We benchmark our approach with the state-of-the-art methods for SR and denoising on the real-world datasets SupER~\cite{SupER} and~\cite{NoiseReal}, respectively.

\begin{figure}
	\centering
	 \includegraphics[width=0.45\textwidth]{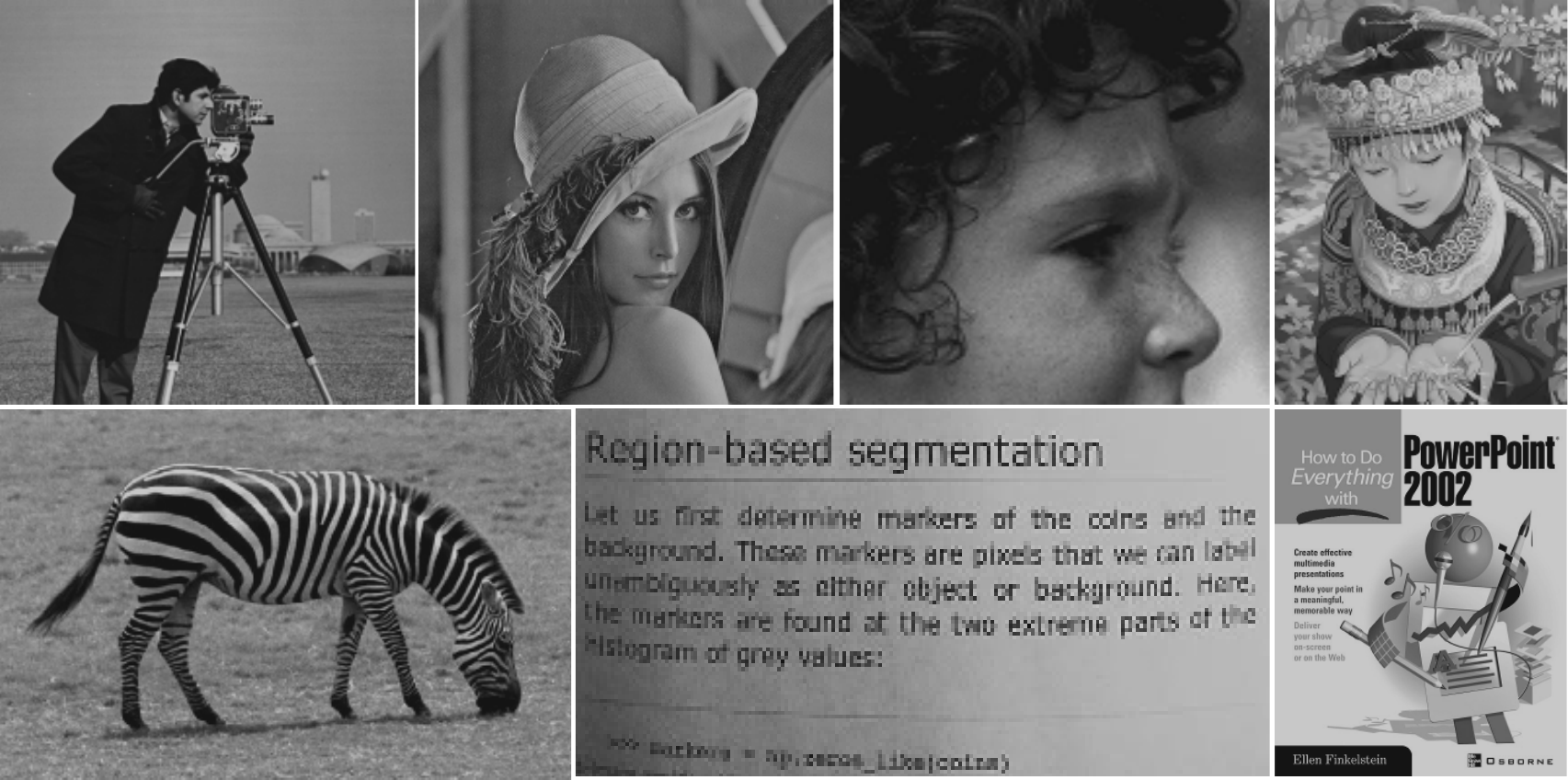}\\
	\caption{8-bit gray-scale natural images for quantitative analysis.}
	\label{fig:SimImgs}
\end{figure}

\begin{figure*}
	\centering
	 \includegraphics[width=0.95\textwidth]{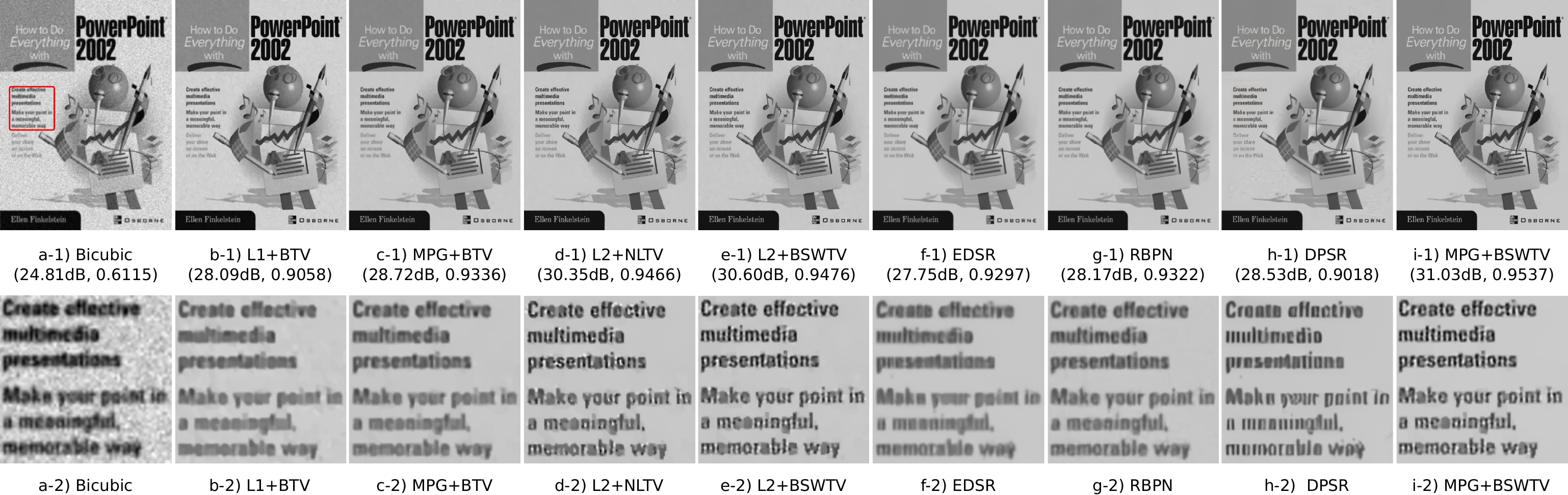}		
	\caption{Comparison of different SR methods for 2$\times$ on $\textit{PPT3}$ contaminated by a mixed Poisson--Gaussian noise with peak intensity 200 and $\sigma = 2$: (a) bicubic, (b) L1+BTV, (c) MPGSR, (d) L2+NLTV, (e) L2+BSWTV, (f) EDSR, (g) RBPN, (h) DPSR, and (i) MPG+BSWTV.}
	\label{fig:SR2}
\end{figure*}

\begin{table*}
\centering
\setlength{\tabcolsep}{2.2pt}
\begin{threeparttable}[b]
\caption{Comparison of different SR methods for 2$\times$ upscaling under a mixed Poisson--Gaussian noise with peak intensity 200, $\sigma = 2$ in PSNR (\textit{\normalfont{dB}}) and SSIM. Best: bold; second best: underline. (All TV-based methods were implemented using ADMM framework.)}
\label{tab:comparisonSR}
%\vspace{-5pt}
	\begin{tabular}{c | c | c | c | c | c | c | c | c}
    \toprule
     & Cameraman & Lena & Page & Comic & Face & PPT3 & Zebra & Average\\
    \midrule
    & PSNR\quad SSIM & PSNR\quad SSIM & PSNR\quad SSIM & PSNR\quad SSIM & PSNR\quad SSIM & PSNR\quad SSIM & PSNR\quad SSIM & PSNR\quad SSIM \\
    \midrule
	Bicubic &26.68\quad 0.6506& 27.27\quad 0.6934&22.80\quad 0.5412&24.31\quad
	0.6921&31.12\quad 0.7271&24.81\quad 0.6115&27.04\quad 0.7224&26.29\quad 0.6626\\
    L1+BTV\cite{sr0} & 29.15\quad 0.8412&29.20\quad 0.8210& 23.75\quad 0.7420&25.96\quad 0.7914&33.40\quad 0.8143& 28.09\quad 0.9058&29.77\quad 0.8064&28.47\quad 0.8174 \\
    L2+NLTV\cite{nltv} &29.96\quad 0.8591& 29.41\quad 0.8277&25.69\quad 0.8133&26.40\quad 0.8196&33.36\quad 0.8080& 30.35\quad 0.9466&30.07\quad 0.8024&29.32\quad 0.8395\\
    MPG+BTV\cite{sun} &29.97\quad 0.8571&\underline{29.84}\quad 0.8410&24.70\quad 0.7864& 26.39\quad 0.8077&33.81\quad 0.8248& 28.72\quad 0.9336&\underline{30.45}\quad \underline{0.8222}&29.13\quad 0.8390\\
    L2+BSWTV & \underline{30.24}\quad \underline{0.8622}&29.75\quad 0.8387&\underline{26.07}\quad \underline{0.8240}&\underline{26.72}\quad \underline{0.8305}&33.67\quad 0.8172&\underline{30.60}\quad \underline{0.9476} &30.42\quad 0.8202&\underline{29.64}\quad \underline{0.8486}\\
    EDSR\cite{EDSR} &28.69\quad 0.8404&28.84\quad 0.8215&23.88\quad 0.7560&24.97\quad 0.7359&33.19\quad 0.7956&27.75\quad 0.9297&28.65\quad 0.7683&28.00\quad 0.8068\\
    RBPN\cite{RBPN} &29.22\quad 0.8593&29.53\quad \underline{0.8458}& 24.67\quad 0.7898&25.99\quad 0.7995&\underline{34.00}\quad \underline{0.8292}&28.17\quad 0.9322&29.54\quad 0.8246&28.73\quad 0.8401\\
    DPSR\cite{DPSR} & 29.08\quad 0.8404&28.72\quad 0.8059&24.59\quad 0.7218& 25.18\quad 0.7461&32.22\quad 0.7652&28.53\quad 0.9018&28.57\quad  0.7729&28.13\quad 0.7934\\
    MPG+BSWTV (ours) &\textbf{30.49}\quad \textbf{0.8706} & \textbf{29.99}\quad \textbf{0.8516}&\textbf{26.26}\quad \textbf{0.8401} &\textbf{27.00}\quad \textbf{0.8360} & \textbf{34.04}\quad \textbf{0.8303} & \textbf{31.03}\quad \textbf{0.9537} & \textbf{30.77}\quad \textbf{0.8465} & \textbf{29.94}\quad \textbf{0.8613} \\
    \bottomrule     
	\end{tabular}
    \end{threeparttable}
\end{table*}

\subsection{Super-Resolution on Synthetic Images}
\label{super-resolution}
We evaluated the proposed MPG+BSWTV on the gray-value reference images shown in Fig.~\ref{fig:SimImgs}. Specially, we assume the system matrix $\mathbf{DBM}$ is known and set the scalar $\alpha=1$. Accordingly, we generated four LR images for each reference image. Firstly, we rescaled the gray-value image to peak intensity 200 and considered as the GT image. The rescaled image was then shifted by $(0, 0), (0.5, 0), (0.5, 0.5)$, and $(0, 0.5)$ pixel to obtain four images. The four images were blurred by an isotropic 3$\times$3 Gaussian kernel and then subsampled by a factor of 2. Each of the degraded LR images was corrupted by a mixed Poisson--Gaussian noise with peak intensity 200 and $\sigma =2$. We compared the proposed method with L1+BTV~\cite{sr0}, L2+NLTV~\cite{nltv}, MPGSR~\cite{sun}, L2+BSWTV, EDSR~\cite{EDSR}, RBPN~\cite{RBPN}, and DPSR~\cite{DPSR}. Note that the notation L2+NLTV indicates the L2-norm data term in conjunction with NLTV as the regularizer and the same notation manner is employed for L1+BTV and L2+BSWTV. To achieve a fair comparision, all the above TV-based methods were implemented by the ADMM algorithm. For a better interpretation, MPGSR is denoted as MPG+BTV in this paper. EDSR and DPSR are CNN-based single-frame SR methods and RBPN is one of the state-of-the-art video SR (VSR) networks which uses multiple LR frames as input. Since EDSR and RBPN are originally trained on noiseless images, we retrained EDSR and RBPN using the original code on the datasets which were contaminated by the mixed Poisson--Gaussian noise with the same noise level as the testing images. For all the investigated methods, the parameters which generated the best PSNR performance were adopted. We demonstrate the reconstructed image $\textit{PPT3}$ by the investigated approaches in Fig.~\ref{fig:SR2}. It is shown that our MPG+BSWTV provides a remarkable improvement quantitatively and qualitatively by jointly enhancing the image resolution and suppressing the residual noise surrounding the characters. The VSR method RBPN generates better result than EDSR as expected by exploiting the information entailed in the neighboring frames. The DPSR tends to suppress the noise aggressively which leads to a degradation of the detailed structures. %Besides, DPSR generates some unpredictable artifacts, for instance, in the zoomed-in region. 
We summarize the quantitative comparison in Tab.~\ref{tab:comparisonSR}.

\begin{figure}
	\centering
	\vspace{-15pt}
	 \includegraphics[width=0.47\textwidth]{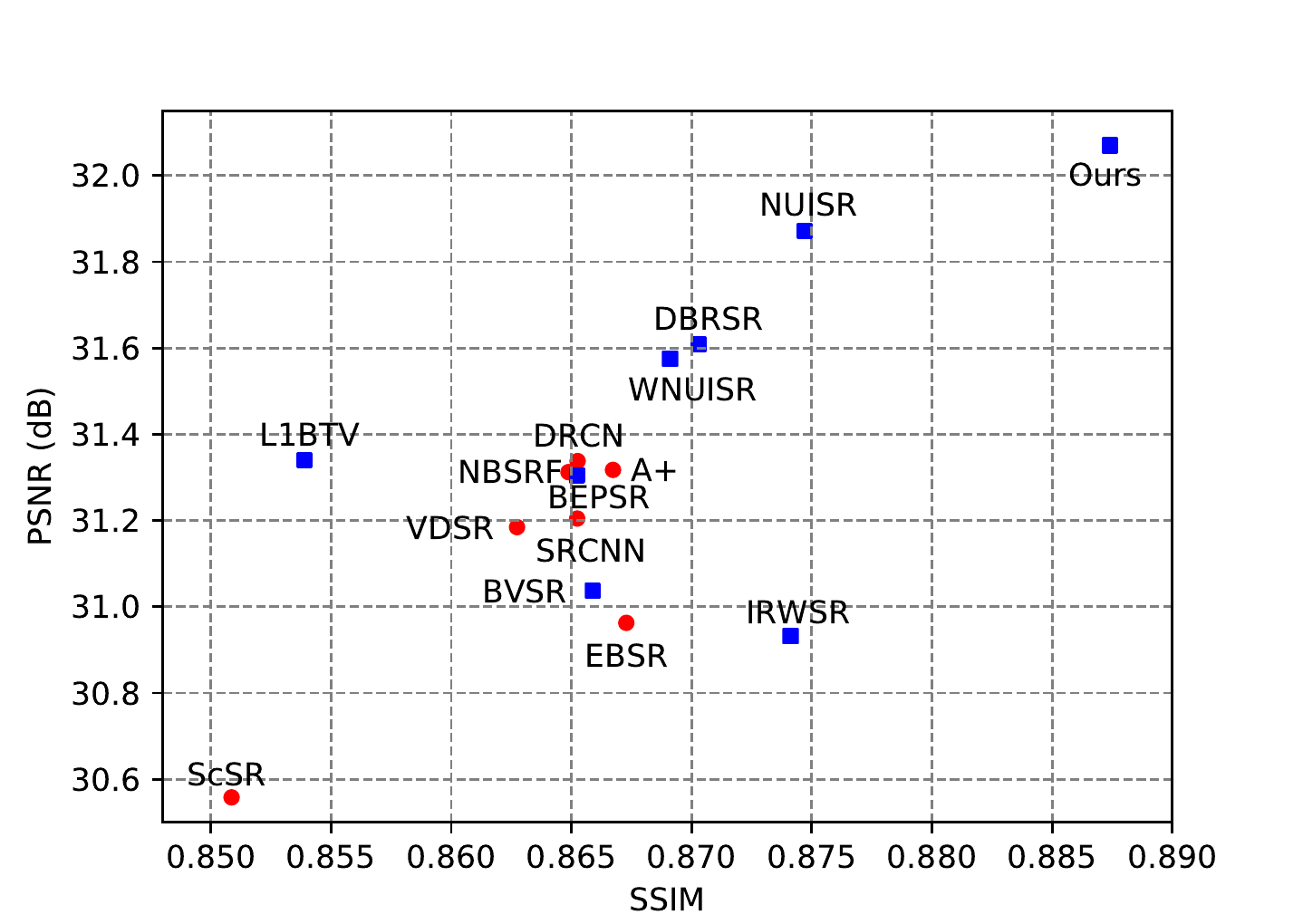}		
	\caption{Comparison with other 14 SR methods on the SupER dataset~\cite{SupER} in average PSNR and SSIM for 2$\times$. Red color map denotes the single-frame SR methods and the blue one represents the multi-frame SR methods.}
	%\vspace{-10pt}
	\label{fig:SupER}
\end{figure}

\begin{figure}
	\centering
	\vspace{-15pt}
	 \includegraphics[width=0.47\textwidth]{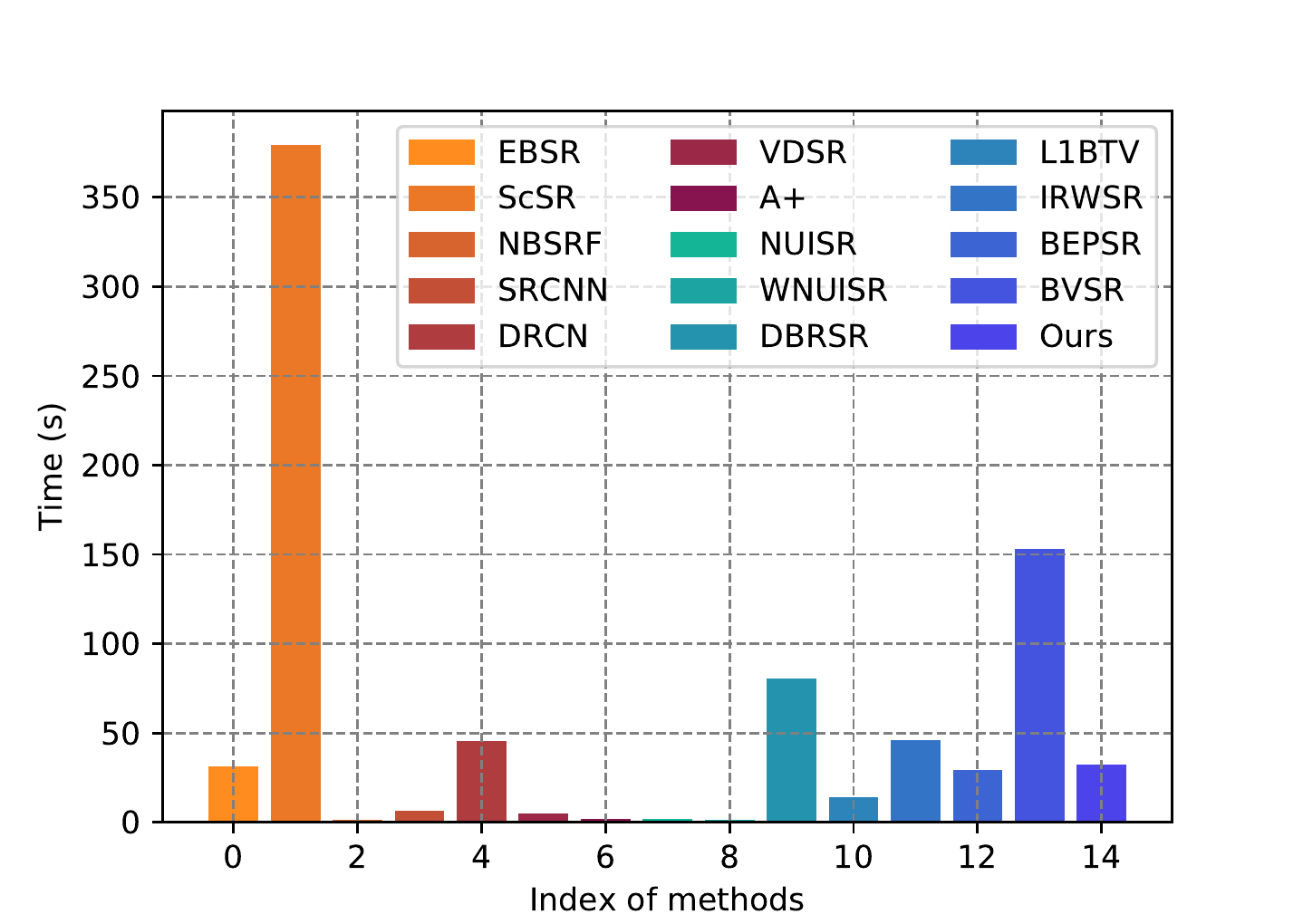}		
	\caption{Comparison with other 14 SR methods on the SupER dataset~\cite{SupER} in runtime for 2$\times$. Red color map denotes the single-frame SR methods and the blue one represents the multi-frame SR methods.}
	%\vspace{-10pt}
	\label{fig:SupERtime}
\end{figure}

\begin{figure*}
	\centering
	 \includegraphics[width=0.95\textwidth]{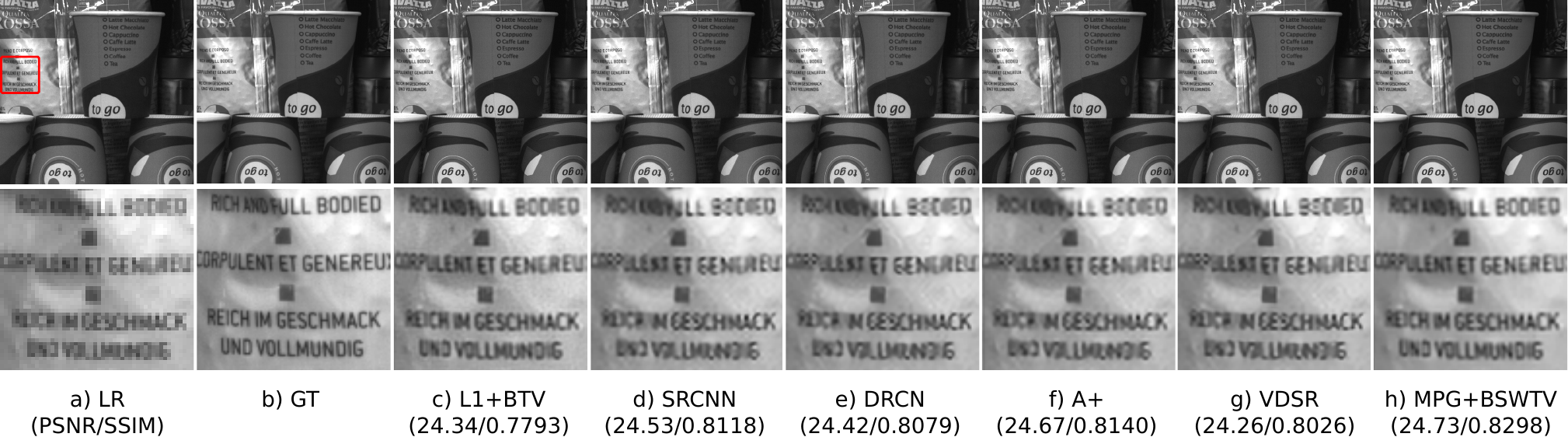}		
	\caption{Comparison of different SR methods on the $\textit{Coffee}$ dataset (2$\times$). Top: reconstructed SR images; Bottom: ROI}
	%\vspace{-10pt}
	\label{fig:bp}
\end{figure*}

\begin{figure*}
	\centering
	 \includegraphics[width=0.95\textwidth]{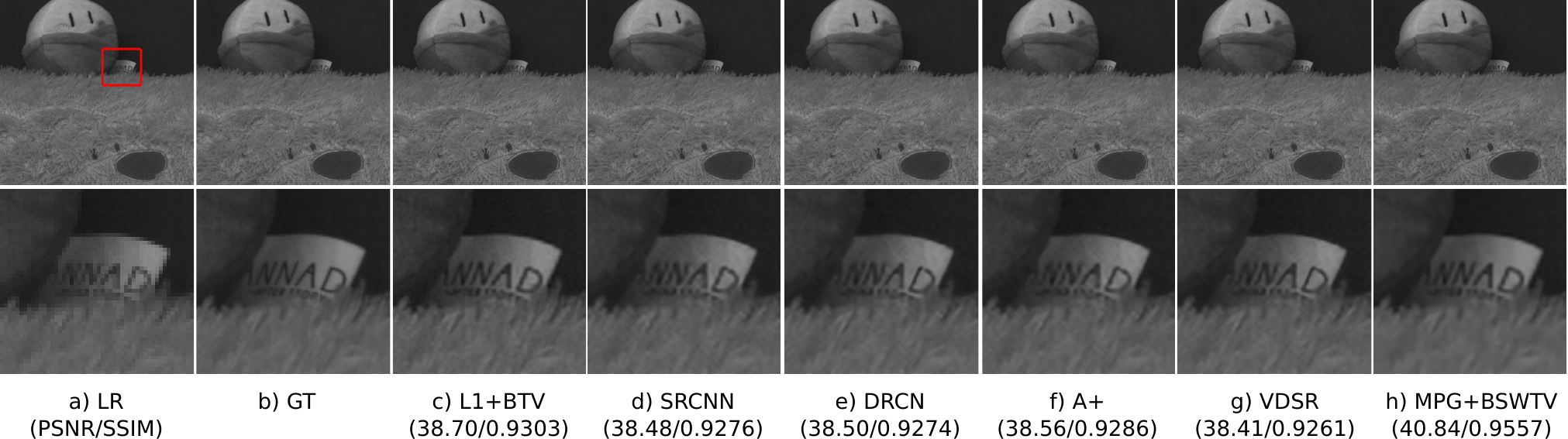}		
	\caption{Comparison of different SR  methods on the $\textit{Dolls}$ dataset (2$\times$). Top: reconstructed SR images; Bottom: ROI}
	%\vspace{-10pt}
	\label{fig:dolls}
\end{figure*}

\subsection{Super-Resolution on Real-World Images}
\label{super-resolution real}

\subsubsection{SR Reconstruction on SupER Dataset}
To validate the proposed method, we conducted experiments on the publicly available SupER dataset~\cite{SupER} which contains images of 14 scenes captured by a CMOS camera. Each of the 14 scenes is captured under multiple modes including motion types, binning factor, and compression levels. Each mode contains 40 LR images by capturing stop-motion videos. We performed SR reconstruction for images captured by binning factor of 2 under global motion which includes translation in 3D space and panning in a joint sinusoidal and circular moving trajectory. Following~\cite{SupER}, we selected a sliding window of size 5 centered at the 10th LR image and compared with other 14 SR methods~\cite{sr0,sr3,kohler,SRCNN, VDSR,DRCN,ScSR,EBSR,NBSRF,A+,WNUISR,DBRSR,Zeng,BVSR} implemented in~\cite{SupER}. Since the LR images are not severely contaminated by noise, we set $\lambda = 0.1$ for all the 14 scenes. We estimated the scalar $\alpha$ and the Gaussian noise parameter $\Ssigma$ shown in Eq.~\eqref{energy function} by~\cite{Egiazarian} and assumed the mean $\Mmu = 0$. The estimated negative parameters by~\cite{Egiazarian} were clamped to $10^{-6}$. Besides, we set the decay scalar as $\gamma = 0.95$ and the penalty parameter as $\rho = 10^3$ for a smooth convergence over 16 iterations. The smoothing parameter $\eta$ was set as 3 to make the flat regions and edges distinguishable in the weighting map. The shift parameter $b$ was tuned as 1 so that the fine structures can be preserved. We used the original implementation and parameters in~\cite{SupER} for the other 14 SR methods. The performance of the 15 SR methods is assessed by PSNR and SSIM and summarized in Fig~\ref{fig:SupER}. Single-frame and multi-frame SR methods are respectively marked by red and blue. We can observe that most of the multi-frame SR methods perform better than the single-frame ones under global motion. The proposed approach achieves considerable improvement comparing to the other investigated methods in both PSNR and SSIM. In Fig.~\ref{fig:SupERtime}, we illustrate the computation time of different methods. It is necessary to note that all the other methods were implemented in Matlab and some of them were accelerated by C++. Our method was implemented in Python without C/C++ speedup. In Fig.~\ref{fig:bp} and Fig.~\ref{fig:dolls}, we demonstrate the reconstructed images of some representative methods. As shown in Fig.~\ref{fig:bp}, comparing to the other methods, the proposed MPG+BSWTV generates more distinguishable characters and much cleaner background. In Fig.~\ref{fig:dolls}, we can observe that our method provides a more pleasant visual perception and resembles the GT image most.

\begin{figure}
	\centering
	 \includegraphics[width=0.45\textwidth]{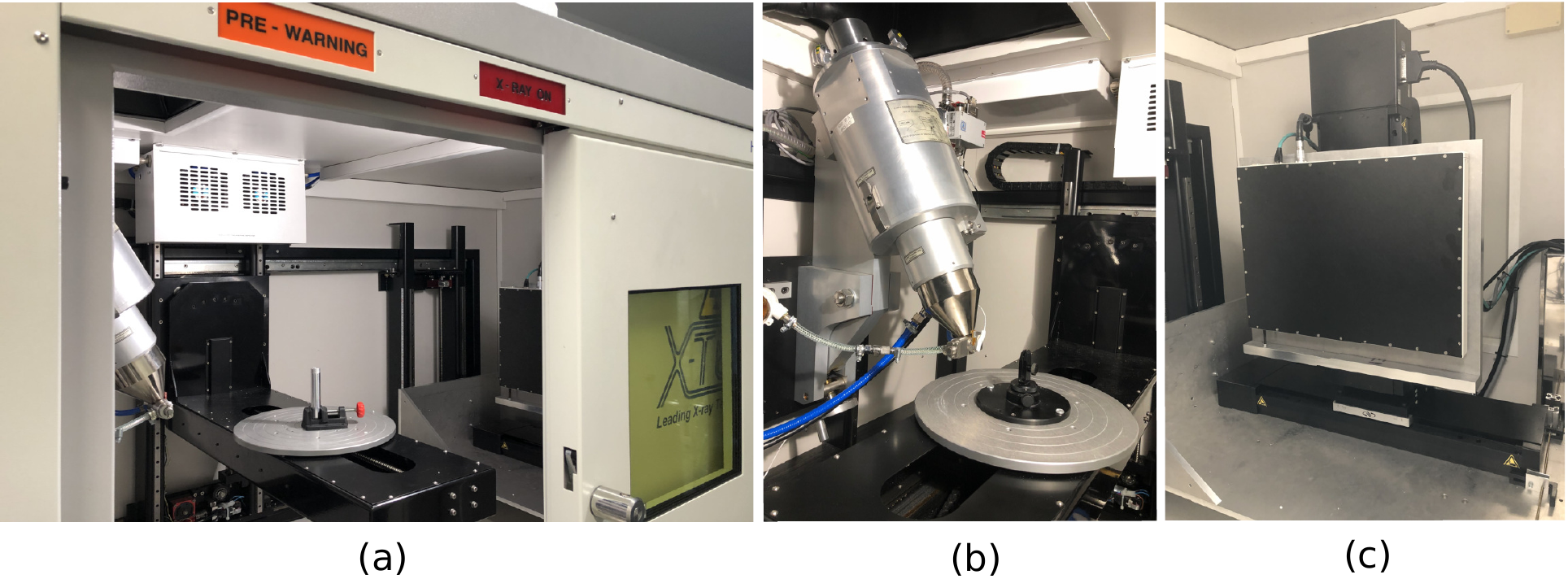}		
	\caption{CT scanner equipped with mounted linear stages. (a) the side view, (b) X-ray source and rotation table, (c) X-ray detector mounted on linear stages.}
	%\vspace{-10pt}
	\label{fig:ct scanner}
\end{figure}

\begin{figure*}
	\centering
	 \includegraphics[width=0.9\textwidth]{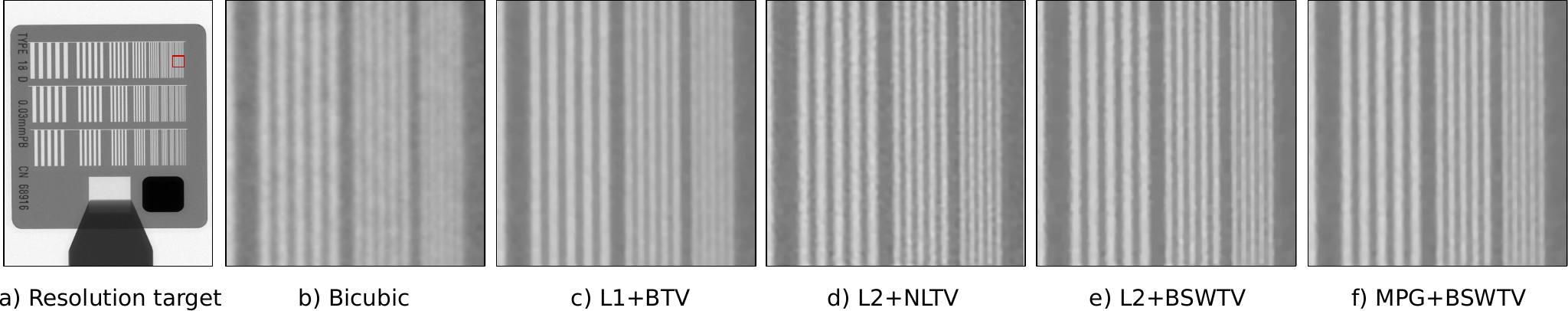}		
	\caption{Comparison of different SR methods for 2$\times$ on the 16-bit X-ray image of a resolution target: (a) X-ray image of the resolution target, (b) bicubic, (c) L1+BTV, (d) L2+NLTV, (e) L2+BSWTV, and (f) MPG+BSWTV.}
	\label{fig:realCT1}
\end{figure*}

\begin{figure*}
	\centering
	 \includegraphics[width=0.9\textwidth]{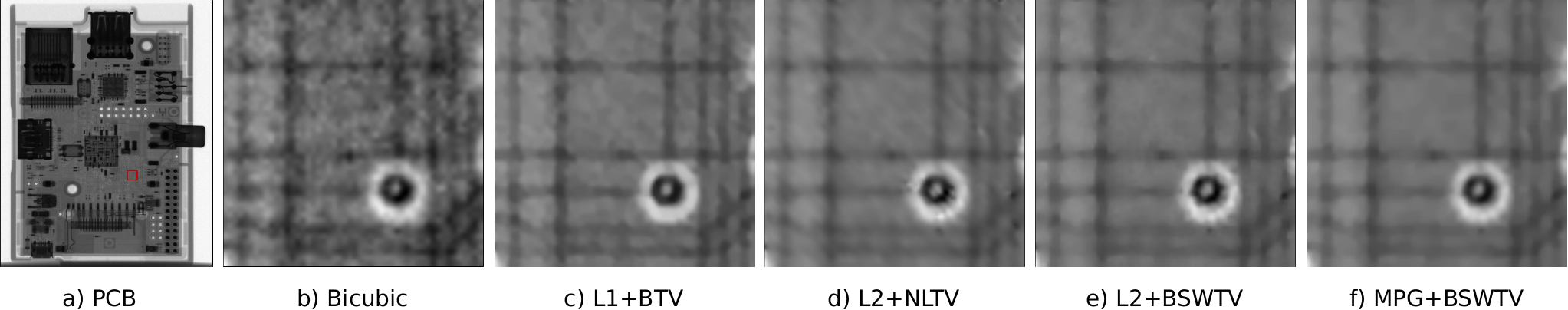}		
	\caption{Comparison of different SR methods for 2$\times$ on the 16-bit X-ray image of a printed circuit board (PCB): (a) X-ray image of the PCB, (b) bicubic, (c) L1+BTV, (d) L2+NLTV, (e) L2+BSWTV, and (f) MPG+BSWTV.}
	\label{fig:realCT3}
\end{figure*}

\subsubsection{SR Reconstruction on X-ray Images}
In addition to the SupER dataset which contains 8-bit natural images, we conducted experiments on 16-bit X-ray images which were captured by the Nikon HMX ST 225 CT scanner as shown in Fig.~\ref{fig:ct scanner}. The CT scanner is equipped with a flat panel Varian PaxScan@4030E detector which has a pixel size of 127$\mu$m$\times$127$\mu$m. The detector is mounted on the controllable linear stages for x- and y-positioning such that the detector can be shifted to a predefined position with a movement accuracy up to 1$\mu$m. Two objects were taken as test specimens: a resolution target and a printed circuit board (PCB). Specially, four 16-bit X-ray images were captured by shifting the detector with a half pixel distance rightwards, downwards, and leftwards for both the specimens. The SR reconstructed images by different methods are demonstrated in Fig.~\ref{fig:realCT1} and Fig.~\ref{fig:realCT3}. It is shown that the proposed method performs better than the others in visual perception by jointly sharpening the edges and suppressing the noise in the flat regions which coincides with the observations in the other SR experiments.  

\subsection{Image Denoising on Synthetic Images}
\label{denoising}

\begin{figure*}
	\centering
	 \includegraphics[width=0.95\textwidth]{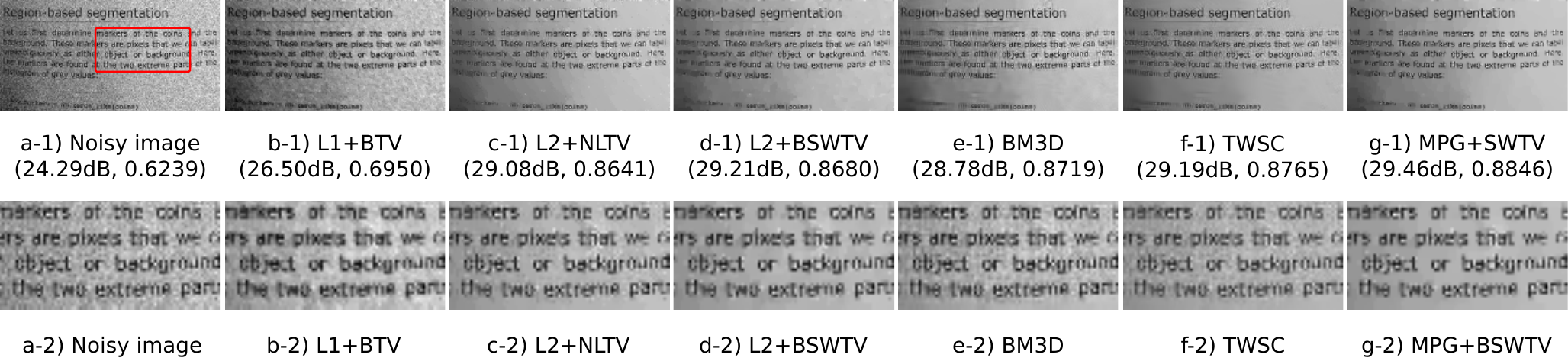}		
	\caption{Comparison of different denoising methods on $\textit{Page}$ contaminated by a mixed Poisson--Gaussian noise with peak intensity 200 and $\sigma = 10$ in PSNR and SSIM: (a) noisy image, (b) L1+BTV, (c) L2+NLTV, (d) L2+BSWTV, (e) BM3D, and (f) MPG+BSWTV.}
	\label{fig:Denoise2}
\end{figure*}

In order to evaluate the performance of the proposed MPG+BSWTV for image denoising, we carried out experiments on the synthetic natural images under mixed Poisson--Gaussian noise. As described in Section~\ref{introduction}, we set the system matrix $A$ as the identity matrix and the amount of input frames as $m = 1$. The experiments were conducted on the gray-value images shown in Fig.~\ref{fig:SimImgs} which were corrupted by a mixed Poisson--Gaussian noise with peak intensity 200 and $\alpha = 0.01, \sigma = 2$. We compared the proposed MPG+BSWTV with L1+BTV~\cite{sr0}, L2+NLTV~\cite{nltv}, L2+BSWTV, BM3D~\cite{BM3D}, and TWSC~\cite{TWSC}. We set the window size as $R =3$ for L2+NLTV and the patch size was selected as $r=3$ for both NLTV and BSWTV. The number of ADMM iterations of all the TV-based methods was set as 20. The decay scalar $\gamma$ is normally tuned in a range of $[0.5, 0.95]$. Empirically, large decay $\gamma$ tends to yield a gradually descent shrink coefficient $\xi$ and results in a flat convergence curve, while small decay usually leads to a fast and slightly fluctuated convergence. The weighting parameter of the regularization term $\lambda$ and the momentum coefficient $\beta$ were tuned to achieve the best PSNR performance in a trial and error manner. The smoothing parameter $\eta$ was selected in a way that the structures were distinguishable in the weighting map. The Gaussian parameter of the kernel $G(\sigma_{\Phi})$ was initialized with $\sigma_{\Phi} = 3$ and the constants for updating the penalty parameters were set as $c_1 = c_2 = 2, c = 10$. The amplitude parameter $a$ in function $f$ shown in Eq.~\eqref{gradient} was chosen as 20 and the shift parameter $b$ was tuned within the interval $[0, 1]$. For the other methods, the parameters which generated the best PSNR were employed. In Fig.~\ref{fig:Denoise2}, we demonstrate the superior performance of our approach for denoising texts. In contrast to BM3D and TWSC, our method restores the characters effectively and meanwhile removes the noise in the empty space which provides a pleasant result. The overall quantitative evaluation in PSNR and SSIM is depicted in Tab~\ref{tab:comparisonDenoising}. 

\begin{table*}
\centering
\setlength{\tabcolsep}{2pt}
\begin{threeparttable}[b]
\caption{Comparison of different denoising methods under mixed Poisson--Gaussian noise with peak intensity 200, $\alpha = 0.01, \sigma = 2$ in PSNR (\textit{\normalfont{dB}}) and SSIM. Best: bold; second best: underline. (All TV-based methods were implemented using ADMM framework.)}
\label{tab:comparisonDenoising}
%\vspace{-0.5pt}
	\begin{tabular}{c | c | c | c | c | c | c | c | c }
    \toprule
     & Cameraman & Lena & Page & Comic & Face & PPT3 & Zebra & Average\\
    \midrule
    & PSNR\quad SSIM & PSNR\quad SSIM & PSNR\quad SSIM & PSNR\quad SSIM & PSNR\quad SSIM & PSNR\quad SSIM & PSNR\quad SSIM & PSNR\quad SSIM\\
    \iffalse
    \multicolumn{9}{c}{Peak intensity 120, $\sigma = 6$}\\
	\midrule
	Noisy image &28.19\quad 0.4180& 28.43\quad 0.4966&27.28\quad 0.5901&27.90\quad
	0.7814&29.24\quad 0.4977&27.00\quad 0.5024&27.99\quad 0.6462 &28.00\quad 0.5618\\
    L1+BTV\cite{sr0} & 33.64\quad 0.7138&33.52\quad 0.7517& 28.39\quad 0.6163&29.34\quad 0.7982&35.46\quad 0.7380& 30.96\quad 0.7803&30.75\quad 0.7120&31.72\quad 0.7300 \\
    L2+NLTV\cite{nltv} &35.75\quad 0.8308&34.68\quad 0.8127&31.29\quad 0.7920&30.48\quad 0.8471&35.76\quad 0.7624& 34.16\quad 0.9225&31.95\quad 0.7692&33.44\quad 0.8195\\
    L2+BSWTV & 36.16\quad \textbf{0.8547}&35.24\quad 0.8234&32.02\quad 0.8276&30.79\quad \underline{0.8610}&36.27\quad \underline{0.7807}&34.50\quad 0.9233 &32.50\quad \underline{0.7894}&33.93\quad 0.8372\\
    BM3D\cite{BM3D} & \textbf{36.74}\quad \underline{0.8541}&\textbf{36.25}\quad \textbf{0.8388}&\underline{32.11}\quad \underline{0.8452}& \textbf{30.99}\quad 0.8608&\textbf{36.66}\quad 0.7686&\textbf{35.11}\quad \underline{0.9328}&\textbf{33.33}\quad 0.7744&\textbf{34.46}\quad \underline{0.8392}\\
    MPG+BSWTV (ours) & \underline{36.53}\quad 0.8485 & \underline{35.71}\quad \underline{0.8353}& \textbf{32.80}\quad \textbf{0.8744} & \textbf{30.99}\quad \textbf{0.8652} & \underline{36.64}\quad \textbf{0.7914} & \underline{34.97}\quad \textbf{0.9391} & \underline{32.93}\quad \textbf{0.7922}& \underline{34.37}\quad \textbf{0.8494} \\
    \fi
       % \multicolumn{9}{c}{Peak intensity 200, $\alpha = 0.01, \sigma = 2$}\\
	\midrule
	Noisy image &40.95\quad 0.9524& 40.93\quad 0.9687&40.71\quad 0.9584&40.72\quad
	0.9928&41.03\quad 0.9711&40.45\quad 0.9555&40.76\quad 0.9856 &40.79\quad 0.9692\\
    L1+BTV\cite{sr0} & 42.25\quad 0.9703&41.19\quad 0.9743& 41.19\quad 0.9737&40.75\quad 0.9932&41.54\quad 0.9747& 41.08\quad 0.9713&40.80\quad 0.9862&41.26\quad 0.9777 \\
    L2+NLTV\cite{nltv} &42.81\quad 0.9752&41.48\quad 0.9756&42.32\quad 0.9817&40.55\quad 0.9934&41.24\quad 0.9734& 44.20\quad 0.9934&41.01\quad 0.9861&41.94\quad 0.9827\\
    L2+BSWTV & 43.43\quad 0.9796&42.08\quad 0.9782&42.96\quad 0.9821&41.22\quad \underline{0.9942}&41.68\quad 0.9756&45.05\quad 0.9952 &41.07\quad \textbf{0.9865}&42.50\quad 0.9845\\
    BM3D\cite{BM3D} & \textbf{43.85}\quad \textbf{0.9833}&41.86\quad \textbf{0.9801}&\underline{43.15}\quad \underline{0.9860}& 41.09\quad 0.9939&41.26\quad \textbf{0.9785}&\underline{45.22}\quad 0.9949&40.99\quad \underline{0.9864}&42.49\quad \textbf{0.9862}\\
    TWSC\cite{TWSC} & \underline{43.81}\quad 0.9815&\textbf{42.23}\quad 0.9786&42.98\quad 0.9833& \underline{41.16}\quad 0.9939&\textbf{42.05}\quad\textbf{0.9785}&\textbf{45.60}\quad \textbf{0.9968}&\textbf{41.28}\quad 0.9863&\textbf{42.73}\quad 0.9856\\
    MPG+BSWTV (ours) & 43.75\quad \underline{0.9826} & \underline{42.21}\quad \underline{0.9788}& \textbf{43.21}\quad \textbf{0.9865} & \textbf{41.28}\quad \textbf{0.9946} & \underline{41.92}\quad \underline{0.9780} & 45.14\quad \underline{0.9962} & \underline{41.17}\quad 0.9863& \underline{42.67}\quad \underline{0.9861} \\
    \bottomrule     
	\end{tabular}
    \end{threeparttable}
\end{table*}

\begin{table*}
\centering
\setlength{\tabcolsep}{2.5pt}
\begin{threeparttable}[b]
\caption{Comparison of different state-of-the-art denoising methods based on 100 cropped real-world images~\cite{NoiseReal} in average PSNR (\textit{\normalfont{dB}}) and SSIM.}
\label{tab:RealDenoising}
%\vspace{-0.5pt}
	\begin{tabular}{c | c | c | c | c | c | c| c | c | c | c | c}
    \toprule
 & CBM3D~\cite{CBM3D} & EPLL~\cite{EPLL} & PGPD~\cite{PGPD} & NCSR~\cite{NCSR} & WNNM~\cite{WNNM} & DnCNN~\cite{DnCNN} & NC~\cite{NC} & CSF~\cite{CSF} & NI~\cite{NI} & TWSC~\cite{TWSC}& MPG+BSWTV\\ \midrule 
 PSNR  & 37.40 & 36.17 & 36.18 &36.40 &36.59 & 36.08 &36.92 & 37.71 & 37.77& \textbf{38.60}&\underline{38.31} \\  \midrule 
 SSIM & 0.9526 & 0.9216 & 0.9206 & 0.9290 &0.9247 & 0.9161 & 0.9449 & 0.9571 & 0.9570 & \textbf{0.9685}& \underline{0.9608} \\
    \bottomrule     
	\end{tabular}
    \end{threeparttable}
\end{table*}

\begin{figure*}[b]
	\centering
	 \includegraphics[width=0.95\textwidth]{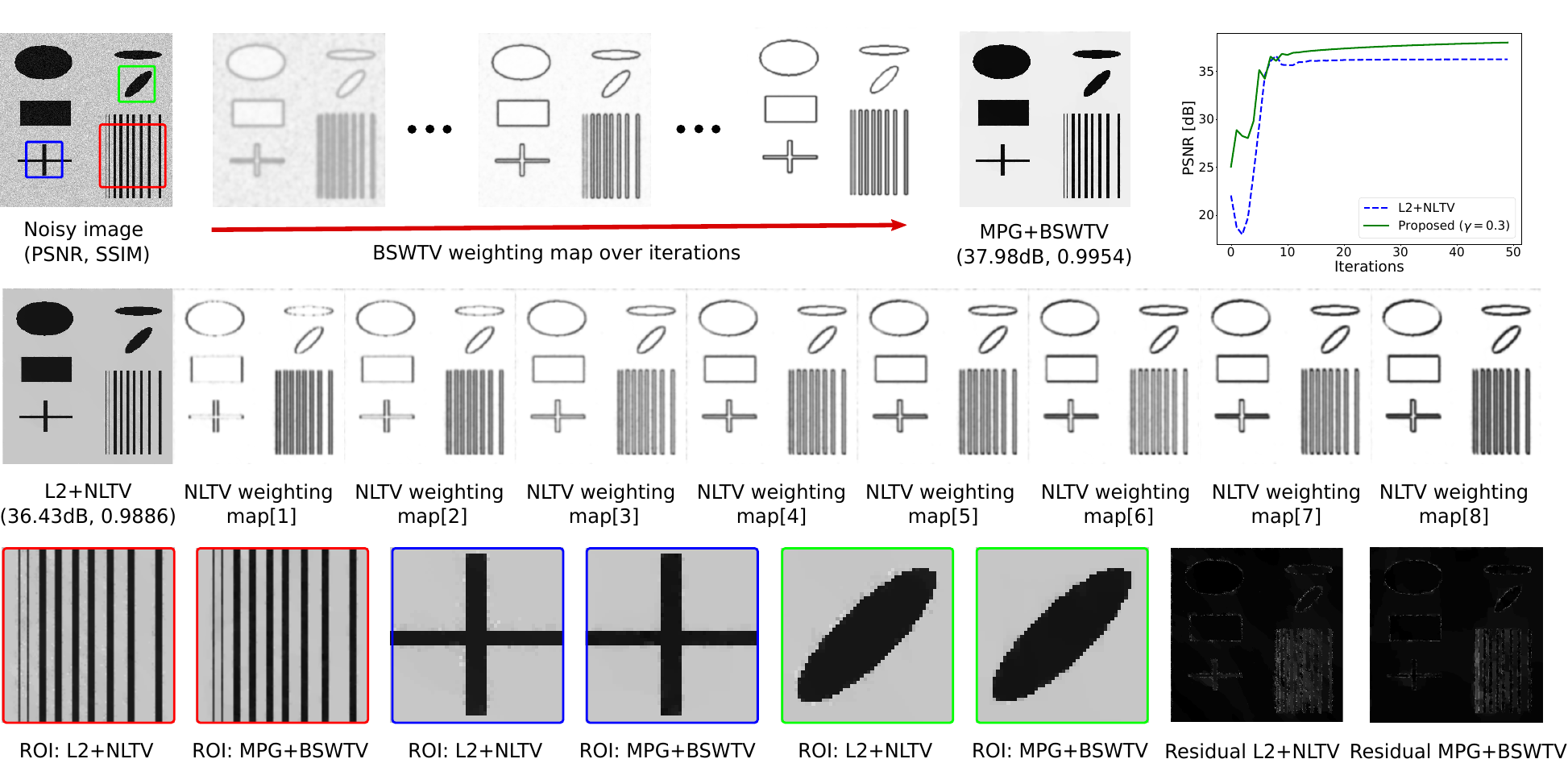}	
	\caption{Illustration of the effectiveness of the shrink coefficient on the weighting map of BSWTV and the reconstructed image comparing to L2+NLTV by denoising an 8-bit gray-value image contaminated by a mixed Poisson--Gaussian noise with peak intensity 200 and $\sigma =10$.}
	\label{fig:AnalysisNoisy}
\end{figure*}

\subsection{Image Denoising on Real-world Images}
\label{Realdenoising}
In this experiment, we evaluated our method on the real-world images~\cite{NoiseReal} for image denoising. The dataset contains noisy images of 40 static scenes which are captured by Canon 5D, Canon 80D, Canon 600D, Nikon D800, and Sony A7 under different ISO levels. 100 cropped regions of size 512$\times$512 are provided in~\cite{NoiseReal}, based on which we compared with other 10 representative denoising methods~\cite{CBM3D,EPLL,PGPD,NCSR,WNNM,CSF,DnCNN,NC,NI,TWSC}. Specially, CBM3D~\cite{CBM3D}, EPLL~\cite{EPLL}, PGPD~\cite{PGPD}, NCSR~\cite{NCSR}, WNNM~\cite{WNNM}, CSF~\cite{CSF}, DnCNN~\cite{DnCNN} are denoising methods based on the AWGN model. As our MPG+BSWTV is proposed for gray-value images, we conducted denoising on each channel of the RGB images individually. The Gaussian noise $\sigma$ and the real scalar $\alpha$ were predicted by~\cite{Egiazarian}. The weighting parameter $\lambda$ was tuned in a range of $[5, 10]$. To obtain a proper weighting map, the smoothing parameter $\eta$ was set as 5. The penalty parameter $\rho$ was initialized with $10^2$ and the number of ADMM iterations was set as 6. Based on the available performance of the other methods in~\cite{NoiseReal}, we summarize the average PSNR and SSIM in Tab.~\ref{tab:RealDenoising}. It is shown that the proposed method MPG+BSWTV performs better than most of the investigated well-known denoising approaches except TWSC which is dedicated to multi-channel images. Besides, we can see that the methods based on AWGN do not perform as well as expected because the real-world noise is not AWGN but signal dependent.

\subsection{Weighting Map and Parameter Analysis}
\label{parameter analysis}
In this section, we illustrate the effectiveness of the shrink coefficient on the weighting map $\PPhi$ and analyze the impact of the penalty parameter $\rho$, the decay scalar $\gamma$, the smoothing parameter $\eta$, and the shift parameter $b$ on the performance of MPG+BSWTV. 

\subsubsection{Weighting map $\PPhi$} In this experiment, we demonstrate the refinement of the weighting map in MPG+BSWTV over ADMM iterations and compare with the resultant weighting maps of L2+NLTV. The synthetic image contains multiple basic shapes including ellipses, rectangles, and bars as shown in Fig.~\ref{fig:AnalysisNoisy}. The image was contaminated by a mixed Poisson--Gaussian noise with peak intensity 200 and $\sigma = 10$. We chose a search window of size $R = 3$ for NLTV and hence NLTV generates $R^2-1$ weighting maps. For a fast convergence, we set the decay parameter as $\gamma =0.3$ and the momentum coefficient as $\beta =0.5$. The smoothing parameters $\eta$ and the weighting parameters $\lambda$ for both NLTV and BSWTV were tuned to achieve the best PSNR performance. As shown in Fig.~\ref{fig:AnalysisNoisy}, both approaches converge over iterations and the proposed method outperforms L2+NLTV quantitatively and qualitatively. We can observe that the mask of the edges in the weighting map of BSWTV becomes thinner and sharper along with the convergence. Consequently, the noise surrounding the edges is significantly suppressed without compromising the sharpness as shown in the marked regions. 

\subsubsection{Penalty parameter $\rho$} 
\begin{figure}
	\centering
	%\vspace{-2pt}
	 \includegraphics[width=0.48\textwidth]{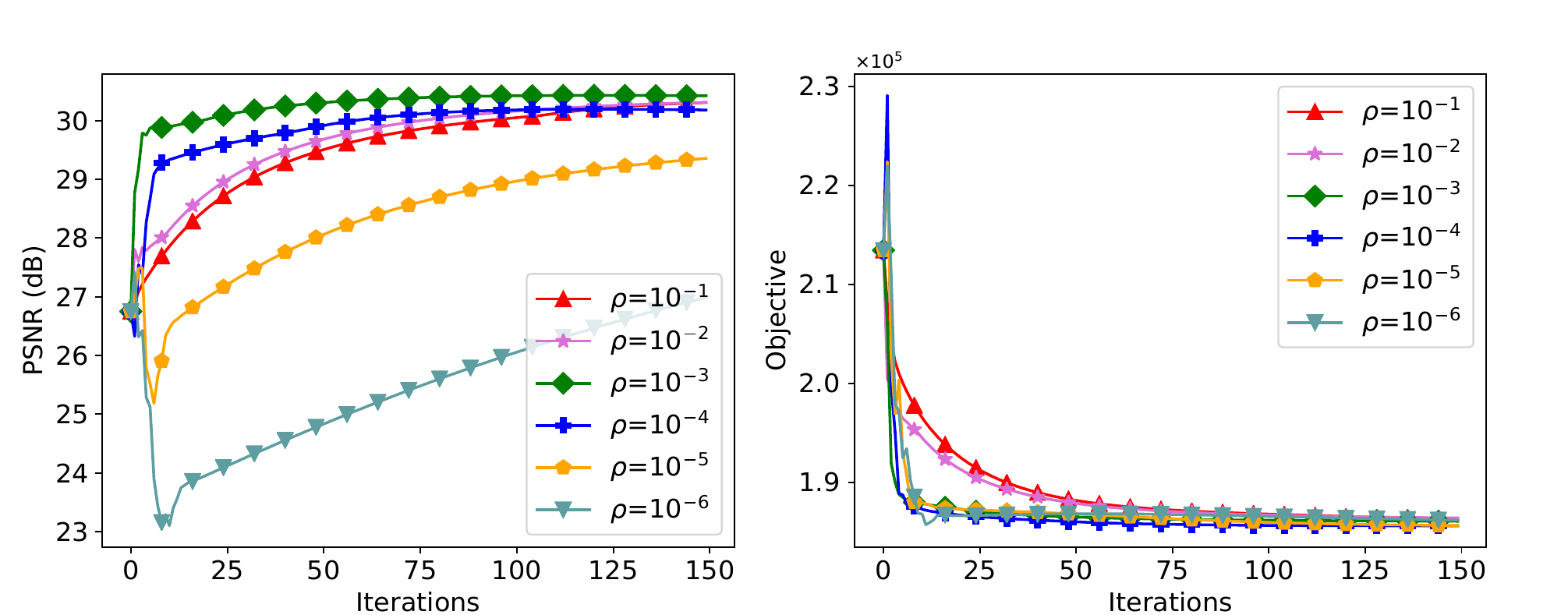}		
	\caption{Impact of the initial penalty parameter $\rho$ on the convergence. Left: PSNR over iterations; Right: objective over iterations.}
	\label{fig:rho}
\end{figure}

Experimental analysis has been conducted to study the influence of different initial $\rho$ on the convergence of the algorithm. As shown in Fig~\ref{fig:rho}, the magnitude of $\rho$ has noticeable impact on the convergence rate although $\rho$ is iteratively updated. Large $\rho$ stabilizes the convergence and tends to slow down the convergence rate. On the contrary, small $\rho$ accelerates the convergence but may cause overshoot of the objective function and lead to undesirable degradation of image quality. Depending on the expected convergence rate and the noise level, an empirical choice of $\rho$ may vary in a range of $[10^{-3},10^{3}]$ for 8-bit images.

\subsubsection{Decay parameter $\gamma$} 
\begin{figure}
	\centering
	%\vspace{-2pt}
	 \includegraphics[width=0.48\textwidth]{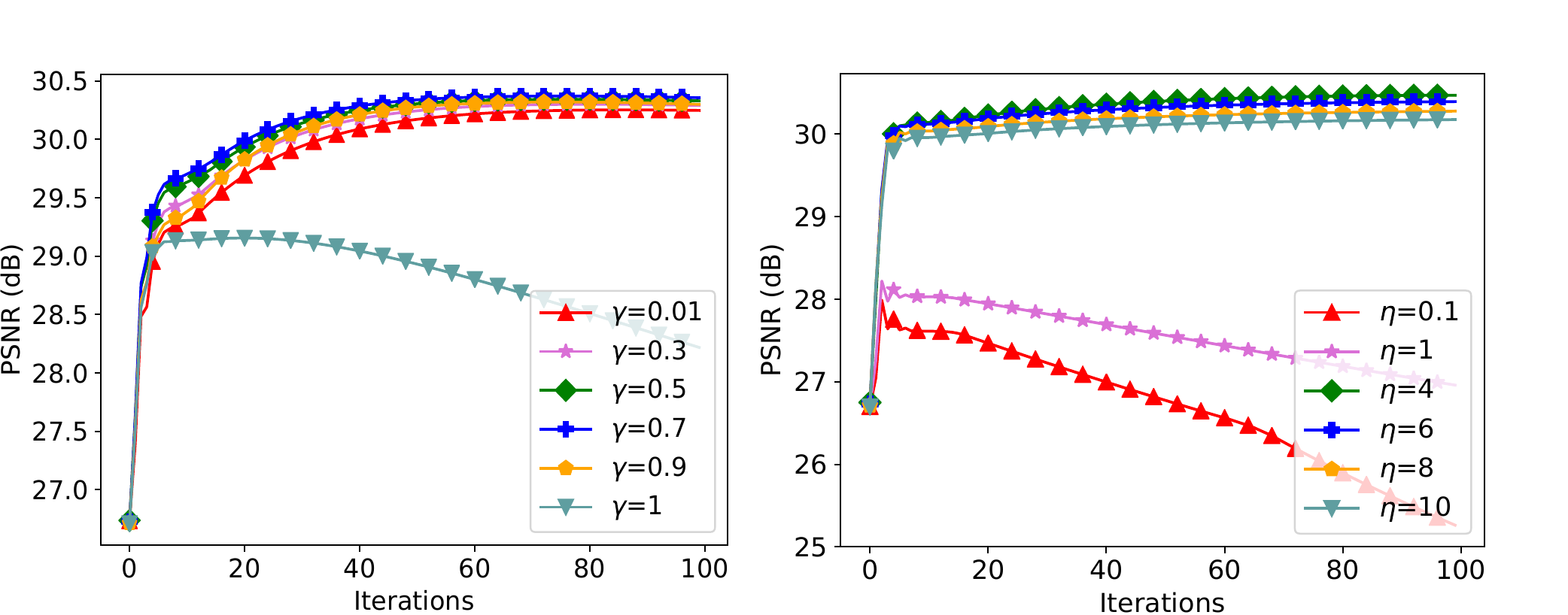}		
	\caption{Left: impact of the decay scalar $\gamma$ on the convergence; Right: impact of the smoothing parameter $\eta$ on the convergence.}
	\label{fig:sigma_eta}
\end{figure}

The decay parameter $\gamma$ aims to refine the weighting map $\Phi$ by thinning the mask of the edges such that the surrounding noise can be better suppressed via TV. We investigated $\gamma$ in the range of $[0.01, 1]$ where $\gamma = 1$ indicates without decay. As shown in the left graph of Fig.~\ref{fig:sigma_eta}, extremely small $\gamma$ attenuates the shrink coefficient $\xi$ aggressively so that the weighting map of regions containing fine low-contrast structures also gets whitened and the fine structures might be smoothed by the regularizer which causes performance degradation. In contrast, $\gamma = 1$ prevents the weighting map from whitening which limits the performance of the algorithm. For a gradual refinement of the weighting map, usually we choose the decay parameter in a range of $[0.5,0.95]$.

\subsubsection{Smoothing parameter $\eta$} The smoothing parameter $\eta$ is used to control the impact of the eigenvalue discrepancy on the weighting map. As depicted in the right graph of Fig.~\ref{fig:sigma_eta}, small $\eta$ can not brighten the weighting map and results in a deteriorated performance. However, too large $\eta$ causes saturation of the weighting map and the proposed regularization term acts as the standard TV. Depending on the dynamic range of the image, a proper choice of $\eta$ for 8-bit images would be in the interval of $[2,6]$. 

\begin{figure}
	\centering
	%\vspace{-2pt}
	 \includegraphics[width=0.48\textwidth]{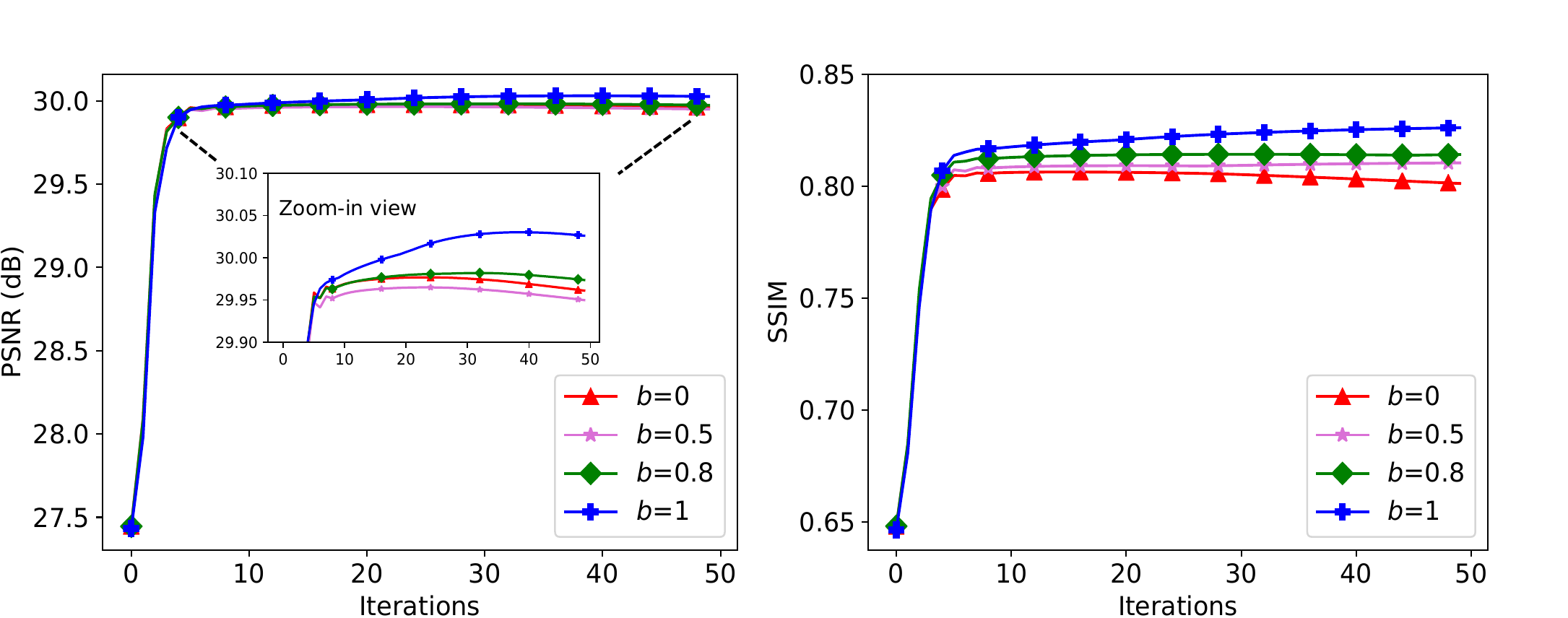}		
	\caption{Impact of the shift parameter $b$ on the convergence. Left: PSNR over iterations; Right: SSIM over iterations.}
	\label{fig:shift}
\end{figure}

\subsubsection{Shift parameter $b$} The flat regions and strong edges can be easily tackled by a homogeneous shrink coefficient. The shift parameter $b$ is introduced to cope with the fine textures with relative low contrast. It behaves as a threshold and masks the fine textures in the weighting map by inhomogeneously shrinking the coefficient $\xi$. Specially, $\xi$ is expected to be shrinked by $\gamma$ in flat regions while maintain the same in fine-structure regions. Consequently, the fine structures are masked out in the weighting map and are not smoothed by TV. As illustrated in Fig.~\ref{fig:shift}, $b$ has noticeable impact on SSIM and might have limited influence on PSNR because PSNR is prone to slightly oversmoothed images.

\section{Conclusion}
\label{Conclusion}
In this paper, we propose a bilateral spectrum weighted total variation (BSWTV) based on the spectrum of the covariance matrix of the adaptively weighted image gradients for real-world noisy-image super-resolution and image denoising. Particularly, we apply a locally adaptive shrink coefficient to the image gradients such that the mask of the edges in the weighting map of the total variation is iteratively refined and the noise surrounding the edges is effectively suppressed. Combining with the data fidelity term MPG derived from a mixed Poisson--Gaussian noise model, we introduce a generalized algorithm addressing real-world non-Gaussian noise for super-resolution and image denoising. The overall objective function is decomposed and solved based on the alternating direction method of multipliers (ADMM) algorithm. Specially, different from the standard ADMM framework, we integrate the update of the weighting map as the first step in the ADMM algorithm by considering the other variables as constants. In order to remove the outliers in the weighting map and facilitate the stability of the convergence of the objective function in the modified ADMM, the estimated weighting map is smoothed by a Gaussian filter with an iteratively decreased kernel width and updated in a momentum-based fashion. We have conducted extensive experiments and benchmarked our approach with the state-of-the-art methods on the publicly available real-world datasets for super-resolution and image denoising. Experimental results demonstrate that our MPG+BSWTV outperforms 14 investigated super-resolution methods by an average gain of 0.2dB in PSNR on the SupER dataset. Although MPG+BSWTV is originally derived for super-resolution, it achieves also promising performance for image denoising on the real-world noisy images. %an average performance gain of 0.35dB in PSNR comparing to the 8 state-of-the-art denoising methods 

\appendices
\section{Proof of Proposition~\ref{App:proof}}
\label{App:proof}

\begin{proof}
Considering a digital image $y: \N^2_0\to \R$ contaminated by a mixed Poisson--Gaussian noise as $y = z + n_p(z) + n_g$ where $(z_{i,j}+n_p(z_{i,j}))/\alpha\sim P(z_{i,j}/\alpha)$ with $z_{i,j}$ being the noiseless intensity value at pixel $(i,j)$ and $\alpha$ being a scalar. $n_g$ is an additive Gaussian noise with $n_g(i,j)\sim N(\mu_{i,j},\sigma^2_{i,j})$. According to the Central Limit Theorem (CLT), when $z_{i,j}/\alpha$ is sufficiently large, we have $(z_{i,j}+n_p(z_{i,j}))/\alpha\sim P(z_{i,j}/\alpha)\simeq N(z_{i,j}/\alpha, z_{i,j}/\alpha)$. Therefore, we have $z_{i,j}+n_p(z_{i,j})\sim N(z_{i,j}, \alpha z_{i,j})$. As $n_p$ and $n_g$ are independent, we yield $y(i,j)\sim N(z_{i,j}+\mu_{i,j}, \alpha z_{i,j}+\sigma^2_{i,j})$. As element $(i+1,j)$ and $(i-1,j)$ are independent, we have $E(\nabla_xy(i,j)) = (z_{i+1,j}+\mu_{i+1,j}-z_{i-1,j}-\mu_{i-1,j})/2$ and $Var(\nabla_xy(i,j)) = (\alpha z_{i+1,j}+\sigma^2_{i+1,j}+\alpha z_{i-1,j}+\sigma^2_{i-1,j})/4$. If elements $(i-1,j), (i,j), (i+1,j)\in \Omega$ where $\forall (m,n), (p,q)$ satisfying $|z_{m,n}+\mu_{m,n}-z_{p,q}-\mu_{p,q}|<\varepsilon_1, |\alpha z_{m,n}+\sigma^2_{m,n}-\alpha z_{p,q}-\sigma^2_{p,q}|<\varepsilon_2, \forall \varepsilon_1, \varepsilon_2>0$, then we have $\nabla_xy(i,j)\sim N(0, (\alpha z_{i,j}+\sigma^2_{i,j})/2)$. The derivation holds also for $\nabla_yy(i,j)$. In the homogeneous region $\Omega$, a collection of the gradients with the same isotropic white Gaussian distribution can be considered as multiple realizations of an isotropic white Gaussian distributed variable at different image locations.  
\end{proof}

\bibliographystyle{elsarticle-num} 
\bibliography{egbib}

\iffalse
\begin{IEEEbiography}{Kaicong Sun}
Biography text here.
\end{IEEEbiography}

\begin{IEEEbiography}{Sven Simon}
Biography text here.
\end{IEEEbiography}
\fi

\end{document}